# Stability of the classical type of relative equilibria of a rigid body in the $J_2$ problem

Yue Wang[*], Shijie Xu

*Room B1024, New Main Building, Department of Guidance, Navigation and Control, School of Astronautics, Beijing University of Aeronautics and Astronautics, 100191, Beijing, China*

**Abstract**

The motion of a point mass in the $J_2$ problem is generalized to that of a rigid body in a $J_2$ gravity field. The linear and nonlinear stability of the classical type of relative equilibria of the rigid body, which have been obtained in our previous paper, are studied in the framework of geometric mechanics with the second-order gravitational potential. Non-canonical Hamiltonian structure of the problem, i.e., Poisson tensor, Casimir functions and equations of motion, are obtained through a Poisson reduction process by means of the symmetry of the problem. The linear system matrix at the relative equilibria is given through the multiplication of the Poisson tensor and Hessian matrix of the variational Lagrangian. Based on the characteristic equation of the linear system matrix, the conditions of linear stability of the relative equilibria are obtained. The conditions of nonlinear stability of the relative equilibria are derived with the energy-Casimir method through the projected Hessian matrix of the variational Lagrangian. With the stability conditions obtained, both the linear and nonlinear stability of the relative equilibria are investigated in details in a wide range of the parameters of the gravity field and the rigid body. We find that both the zonal harmonic $J_2$ and the characteristic dimension of the rigid body have significant effects on the linear and nonlinear stability. Similar to the classical attitude stability in a central gravity field, the linear stability region is also consisted of two regions that are analogues of the Lagrange region and the DeBra-Delp region respectively. The nonlinear stability region is the subset of the linear stability region in the first quadrant that is the analogue of the Lagrange region. Our results are very useful for the studies on the motion of natural satellites in our solar system.

[*]Corresponding author. Tel.: +86 10 8233 9751.
  *E-mail addresses:* ywang@sa.buaa.edu.cn (Y. Wang), starsjxu@yahoo.com.cn (S. Xu).

## 1. Introduction

The $J_2$ problem, also called main problem of artificial satellite theory, in which the motion of a point mass in a gravity field truncated on the zonal harmonic $J_2$ is studied, is an important problem in the celestial mechanics and astrodynamics (Broucke 1994). The $J_2$ problem has its wide applications in the orbital dynamics and orbital design of spacecraft. This classical problem has been studied by many authors, such as Broucke (1994) and the literatures cited therein.

However, neither natural nor artificial celestial bodies are point masses or have spherical mass distributions. One of the generalizations of the point mass model is the rigid body model. Because of the non-spherical mass distribution, the orbital and rotational motions of the rigid body are coupled through the gravity field. The orbit-rotation coupling may cause qualitative effects on the motion, which are more significant when the ratio of the dimension of rigid body to the orbit radius is larger.

The orbit-rotation coupling and its qualitative effects have been discussed in several works on the motion of a rigid body or gyrostat in a central gravity field (Wang et al. 1991, 1992, 1995; Teixidó Román 2010). In Wang and Xu (in press), the orbit-rotation coupling of a rigid satellite around a spheroid planet was assessed. It was found that the significant orbit-rotation coupling should be considered for a spacecraft orbiting a small asteroid or an irregular natural satellite around a planet.

The effects of the orbit-rotation coupling have also been considered in many works on the Full Two Body Problem (F2BP), the problem of the rotational and orbital motions of two rigid bodies interacting through their mutual gravitational potential. A spherically-simplified model of F2BP, in which one body is assumed to be a homogeneous sphere, has been studied broadly, such as Kinoshita (1970), Barkin (1979), Aboelnaga and Barkin (1979), Beletskii and Ponomareva (1990), Scheeres (2004), Breiter et al. (2005), Balsas et al. (2008), Bellerose and Scheeres (2008) and Vereshchagin et al. (2010). There are also several works on the more general models of F2BP, in which both bodies are non-spherical, such as Maciejewski (1995), Scheeres (2002, 2009), Koon et al. (2004), Boué and Laskar (2009) and McMahon and Scheeres (in press).

When the dimension of the rigid body is very small in comparison with the orbital radius, the orbit-rotation coupling is not significant. In the case of an artificial Earth satellite, the point mass model of the $J_2$ problem works very well. However, when a spacecraft orbiting around an asteroid or an irregular natural satellite around a planet, such as Phobos, is considered, the mass distribution of the considered body is far from a sphere and the dimension of the body is not small anymore in comparison with the orbital radius. In these cases, the orbit-rotation coupling causes significant effects and should be taken into account in the precise theories of the motion, as shown by Koon et al. (2004), Scheeres (2006), Wang and Xu (in press).

For the high-precision applications in the coupled orbital and rotational motions of a spacecraft orbiting a spheroid asteroid, or an irregular natural satellite around a dwarf planet or planet, we have generalized the $J_2$ problem to the motion of a rigid body in a $J_2$ gravity field in our previous paper (Wang and Xu 2013a). In that paper, the relative equilibria of the rigid body were determined from a global point of view in the framework of geometric mechanics. A classical type of relative equilibria, as well as a non-classical type of relative equilibria, was uncovered under the second-order gravitational potential.

Through the non-canonical Hamiltonian structure of the problem, geometric mechanics provides a systemic and effective method for determining the linear and nonlinear stability of the relative equilibria, as shown by Beck and Hall (1998). The linear and nonlinear stability of the classical type of relative equilibria already obtained in Wang and Xu (2013a) will be studied further in this paper in the framework of geometric mechanics. Through the stability properties of the relative equilibria, it is sufficient to understand the general dynamical properties of the system near the relative equilibria to a big extent.

Notice that the problem in McMahon and Scheeres (in press) is very similar to our problem. In their paper, the existence of stable equilibrium points, and the linearized and nonlinear dynamics around equilibrium points in the planar F2BP with an oblate primary body were investigated. The differences with our problem are that in their problem the motion is restricted on the equatorial plane of the primary body and the mass center of the primary body is not fixed in the inertial space.

The equilibrium configuration exists generally among the natural celestial bodies in our solar system. It is well known that many natural satellites of big planets evolved tidally to the state of synchronous motion (Wisdom 1987). Notice that the gravity field of the big planets can be well approximated by a $J_2$ gravity field. The results on the stability of the relative equilibria in our problem are very useful for the studies on the motion of many natural satellites.

We also make comparisons with previous results on the stability of the relative equilibria of a rigid body in a central gravity field, such as Wang et al. (1991) and Teixidó Román (2010). The influence of the zonal harmonic $J_2$ on the stability of the relative equilibria is discussed in details.

## 2. Non-canonical Hamiltonian Structure and Relative Equilibria

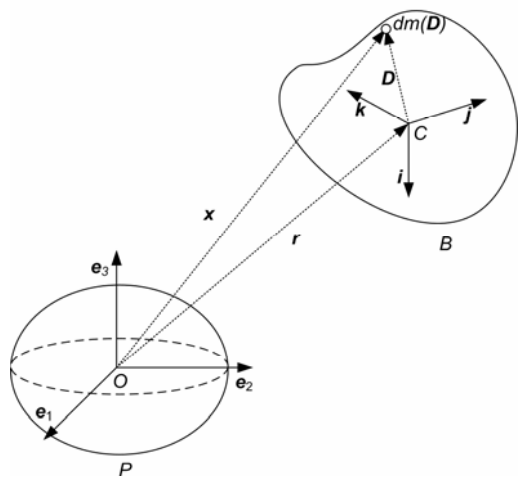

**Fig. 1.** A small rigid body $B$ in the $J_2$ gravity field of a massive axis-symmetrical body $P$

The problem we studied here is same as in Wang and Xu (2013a). As described in Fig. 1, we consider a small rigid body *B* in the gravity field of a massive axis-symmetrical body *P*. Assume that *P* is rotating uniformly around its axis of symmetry, and the mass center of *P* is stationary in the inertial space, i.e. *P* is in free motion without being affected by *B*. The gravity field of *P* is approximated through truncation on the second zonal harmonic $J_2$. The inertial reference frame is defined as $S=\{e_1, e_2, e_3\}$ with its origin *O* attached to the mass center of *P*. $e_3$ is along the axis of symmetry of *P*. The body-fixed reference frame is defined as $S_b=\{i, j, k\}$ with its origin *C* attached to the mass center of *B*. The frame $S_b$ coincides with the principal axes reference frame of *B*.

In Wang and Xu (2013a), a Poisson reduction was applied on the original system by means of the symmetry of the problem. After the reduction process, the non-canonical Hamiltonian structure, i.e., Poisson tensor, Casimir functions and equations of motion, and a classical kind of relative equilibria of the problem were obtained. Here we only give the basic description of the problem and list the main results obtained by us there, see that paper for the details.

The attitude matrix of the rigid body *B* with respect to the inertial frame *S* is denoted by *A*,

$$A = [i, j, k] \in SO(3), \qquad (1)$$

where the vectors *i*, *j* and *k* are expressed in the frame *S*, and *SO*(3) is the 3-dimensional special orthogonal group. *A* is the coordinate transformation matrix from the frame $S_b$ to the frame *S*. If $W = [W^x, W^y, W^z]^T$ are components of a vector in frame $S_b$, its components in frame $S$ can be calculated by

$$w = AW. \tag{2}$$

We define $r$ as the radius vector of point $C$ with respect to $O$ in frame $S$. The radius vector of a mass element $dm(D)$ of the body $B$ with respect to $C$ in frame $S_b$ is denoted by $D$, then the radius vector of $dm(D)$ with respect to $O$ in frame $S$, denoted by $x$, is

$$x = r + AD. \tag{3}$$

Therefore, the configuration space of the problem is the Lie group

$$Q = SE(3), \tag{4}$$

known as the Euclidean group of three space with elements $(A, r)$ that is the semidirect product of $SO(3)$ and $\mathbb{R}^3$. The elements $\Xi$ of the phase space, the cotangent bundle $T^*Q$, can be written in the following coordinates

$$\Xi = (A, r; A\hat{\Pi}, p), \tag{5}$$

where $\Pi$ is the angular momentum expressed in the body-fixed frame $S_b$ and $p$ is the linear momentum of the rigid body expressed in the inertial frame $S$ (Wang and Xu 2012). The hat map $\wedge : \mathbb{R}^3 \to so(3)$ is the usual Lie algebra isomorphism, where $so(3)$ is the Lie Algebras of Lie group $SO(3)$.

The phase space $T^*Q$ carries a natural symplectic structure $\omega = \omega^{SE(3)}$, and the canonical bracket associated to $\omega$ can be written in coordinates $\Xi$ as

$$\{f, g\}_{T^*Q}(\Xi) = \langle D_A f, D_{A\hat{\Pi}} g \rangle - \langle D_A g, D_{A\hat{\Pi}} f \rangle + \left(\frac{\partial f}{\partial r}\right)^T \frac{\partial g}{\partial p} - \left(\frac{\partial g}{\partial r}\right)^T \frac{\partial f}{\partial p}, \tag{6}$$

for any $f, g \in C^\infty(T^*Q)$, $\langle \cdot, \cdot \rangle$ is the pairing between $T^*SO(3)$ and $TSO(3)$, and $D_B f$ is a matrix whose elements are the partial derivates of the function $f$ with respect to the elements of matrix $\boldsymbol{B}$ respectively (Wang and Xu 2012).

The Hamiltonian of the problem $H: T^*Q \to \mathbb{R}$ is given as follows

$$H = \frac{|\boldsymbol{p}|^2}{2m} + \frac{1}{2}\boldsymbol{\Pi}^T \boldsymbol{I}^{-1} \boldsymbol{\Pi} + V \circ \tau_{T*Q}, \tag{7}$$

where $m$ is the mass of the rigid body, the matrix $\boldsymbol{I} = diag\{I_{xx}, I_{yy}, I_{zz}\}$ is the tensor of inertia of the rigid body and $\tau_{T*Q}: T^*Q \to Q$ is the canonical projection.

According to Wang and Xu (2013a), the gravitational potential $V: Q \to \mathbb{R}$ up to the second order is given in terms of moments of inertia as follows:

$$V = V^{(0)} + V^{(2)} = -\frac{GM_1 m}{R} - \frac{GM_1}{2R^3}\left[tr(\boldsymbol{I}) - 3\bar{\boldsymbol{R}}^T \boldsymbol{I} \bar{\boldsymbol{R}} + \varepsilon m - 3\varepsilon m(\boldsymbol{\gamma} \cdot \bar{\boldsymbol{R}})^2\right], \tag{8}$$

where $G$ is the Gravitational Constant, and $M_1$ is the mass of the body $P$. The parameter $\varepsilon$ is defined as $\varepsilon = J_2 a_E^2$, where $a_E$ is the mean equatorial radius of $P$. $\boldsymbol{\gamma}$ is the unit vector $\boldsymbol{e}_3$ expressed in the frame $S_b$. $\boldsymbol{R} = A^T \boldsymbol{r}$ is the radius vector of the mass center of $B$ expressed in frame $S_b$. Note that $R = |\boldsymbol{R}|$ and $\bar{\boldsymbol{R}} = \boldsymbol{R}/R$.

The $J_2$ gravity field is axis-symmetrical with axis of symmetry $\boldsymbol{e}_3$. According to Wang and Xu (2012), the Hamiltonian of the system is $S^1$-invariant, namely the system has symmetry, where $S^1$ is the one-sphere. Using this symmetry, we have carried out a reduction, induced a Hamiltonian on the quotient $T^*Q/S^1$, and expressed the dynamics in terms of appropriate reduced variables in Wang and Xu (2012), where $T^*Q/S^1$ is the quotient of the phase space $T^*Q$ with respect to the action of $S^1$. The reduced variables in $T^*Q/S^1$ can be chosen as

$$\boldsymbol{z} = \left[\boldsymbol{\Pi}^T, \boldsymbol{\gamma}^T, \boldsymbol{R}^T, \boldsymbol{P}^T\right]^T \in \mathbb{R}^{12}, \tag{9}$$

where $\boldsymbol{P} = A^T \boldsymbol{p}$ is the linear momentum of the body $B$ expressed in the body-fixed frame $S_b$ (Wang and Xu 2012). The projection from $T^*Q$ to $T^*Q/S^1$ is given by

$$\Psi\left(\boldsymbol{A},\boldsymbol{r};\boldsymbol{A}\hat{\boldsymbol{\Pi}},\boldsymbol{p}\right)=\left[\boldsymbol{\Pi}^{T},\boldsymbol{\gamma}^{T},\boldsymbol{R}^{T},\boldsymbol{P}^{T}\right]^{T}. \tag{10}$$

According to Marsden and Ratiu (1999), there is a unique non-canonical Hamiltonian structure on $T^*Q/S^1$ such that $\Psi$ is a Poisson map. That is to say, there is a unique Poisson bracket $\{\cdot,\cdot\}_{\mathbb{R}^{12}}(z)$ such that

$$\{f,g\}_{\mathbb{R}^{12}}(z)\circ\Psi=\{f\circ\Psi,g\circ\Psi\}_{T^*Q}(\boldsymbol{\Xi}), \tag{11}$$

for any $f,g\in C^{\infty}(\mathbb{R}^{12})$, where $\{\cdot,\cdot\}_{T^*Q}(\boldsymbol{\Xi})$ is the natural canonical bracket of the system given by Eq. (6).

According to Wang and Xu (2012), the Poisson bracket $\{\cdot,\cdot\}_{\mathbb{R}^{12}}(z)$ can be written in the following form

$$\{f,g\}_{\mathbb{R}^{12}}(z)=\left(\nabla_z f\right)^T \boldsymbol{B}(z)\left(\nabla_z g\right), \tag{12}$$

with the Poisson tensor $\boldsymbol{B}(z)$ given by

$$\boldsymbol{B}(z)=\begin{bmatrix}\hat{\boldsymbol{\Pi}} & \hat{\boldsymbol{\gamma}} & \hat{\boldsymbol{R}} & \hat{\boldsymbol{P}}\\ \hat{\boldsymbol{\gamma}} & 0 & 0 & 0\\ \hat{\boldsymbol{R}} & 0 & 0 & \boldsymbol{E}\\ \hat{\boldsymbol{P}} & 0 & -\boldsymbol{E} & 0\end{bmatrix}, \tag{13}$$

where $\boldsymbol{E}$ is the identity matrix. This Poisson tensor has two independent Casimir functions. One is a geometric integral $C_1(z)=\frac{1}{2}\boldsymbol{\gamma}^T\boldsymbol{\gamma}\equiv\frac{1}{2}$, and the other one is $C_2(z)=\boldsymbol{\gamma}^T\left(\boldsymbol{\Pi}+\hat{\boldsymbol{R}}\boldsymbol{P}\right)$, the third component of the angular momentum with respect to origin $O$ expressed in the inertial frame $S$. $C_2(z)$ is the conservative quantity produced by the symmetry of the system, as stated by Noether's theorem.

The ten-dimensional invariant manifold or symplectic leaf of the system is defined in $\mathbb{R}^{12}$ by Casimir functions

$$\Sigma=\left\{\left(\boldsymbol{\Pi}^T,\boldsymbol{\gamma}^T,\boldsymbol{R}^T,\boldsymbol{P}^T\right)^T\in\mathbb{R}^{12}\mid\boldsymbol{\gamma}^T\boldsymbol{\gamma}=1,\boldsymbol{\gamma}^T\left(\boldsymbol{\Pi}+\hat{\boldsymbol{R}}\boldsymbol{P}\right)=\text{constant}\right\}, \tag{14}$$

which is actually the reduced phase space $T^*\left(Q/S^1\right)$ of the symplectic reduction.

The restriction of the Poisson bracket $\{\cdot,\cdot\}_{\mathbb{R}^{12}}(z)$ to $\Sigma$ defines the symplectic structure on this symplectic leaf.

The equations of motion of the system can be written in the Hamiltonian form

$$\dot{z} = \{z, H(z)\}_{\mathbb{R}^{12}}(z) = \boldsymbol{B}(z)\nabla_z H(z). \tag{15}$$

With the Hamiltonian $H(z)$ given by Eq. (7), the explicit equations of motion are given by

$$\begin{aligned}
\dot{\boldsymbol{\Pi}} &= \boldsymbol{\Pi} \times \boldsymbol{I}^{-1}\boldsymbol{\Pi} + \boldsymbol{R} \times \frac{\partial V(\boldsymbol{\gamma}, \boldsymbol{R})}{\partial \boldsymbol{R}} + \boldsymbol{\gamma} \times \frac{\partial V(\boldsymbol{\gamma}, \boldsymbol{R})}{\partial \boldsymbol{\gamma}}, \\
\dot{\boldsymbol{\gamma}} &= \boldsymbol{\gamma} \times \boldsymbol{I}^{-1}\boldsymbol{\Pi}, \\
\dot{\boldsymbol{R}} &= \boldsymbol{R} \times \boldsymbol{I}^{-1}\boldsymbol{\Pi} + \frac{\boldsymbol{P}}{m}, \\
\dot{\boldsymbol{P}} &= \boldsymbol{P} \times \boldsymbol{I}^{-1}\boldsymbol{\Pi} - \frac{\partial V(\boldsymbol{\gamma}, \boldsymbol{R})}{\partial \boldsymbol{R}}.
\end{aligned} \tag{16}$$

Based on the equations of motion Eq. (16), we have obtained a classical kind of relative equilibria of the rigid body under the second-order gravitational potential in Wang and Xu (2013a). At this type of relative equilibria, the orbit of the mass center of the rigid body is a circle in the equatorial plane of body $P$ with its center coinciding with origin $O$. The rigid body rotates uniformly around one of its principal axes that is parallel to $\boldsymbol{e}_3$ in the inertial frame $S$ in angular velocity that is equal to the orbital angular velocity $\boldsymbol{\Omega}_e$. The radius vector $\boldsymbol{R}_e$ and the linear momentum $\boldsymbol{P}_e$ are parallel to another two principal axes of the rigid body.

When the radius vector $\boldsymbol{R}_e$ is parallel to the principal axes of the rigid body $\boldsymbol{i}, \boldsymbol{j}, \boldsymbol{k}$, the norm of the orbital angular velocity $\boldsymbol{\Omega}_e$ is given by the following three equations respectively:

$$\Omega_e = \left( \frac{GM_1}{R_e^3} + \frac{3GM_1}{2R_e^5}\left[ -2\frac{I_{xx}}{m} + \frac{I_{yy}}{m} + \frac{I_{zz}}{m} + \varepsilon \right] \right)^{1/2}, \tag{17}$$

$$\Omega_e = \left( \frac{GM_1}{R_e^3} + \frac{3GM_1}{2R_e^5} \left[ \frac{I_{xx}}{m} - 2\frac{I_{yy}}{m} + \frac{I_{zz}}{m} + \varepsilon \right] \right)^{1/2}, \tag{18}$$

$$\Omega_e = \left( \frac{GM_1}{R_e^3} + \frac{3GM_1}{2R_e^5} \left[ \frac{I_{xx}}{m} + \frac{I_{yy}}{m} - 2\frac{I_{zz}}{m} + \varepsilon \right] \right)^{1/2}. \tag{19}$$

The norm of the linear momentum $P_e$ is given by:

$$P_e = mR_e\Omega_e. \tag{20}$$

With a given value of $R_e$, there are 24 relative equilibria belonging to this classical type in total. Without of loss of generality, we will choose one of the relative equilibria as shown by Fig. 2 for stability conditions

$$\begin{aligned}
\boldsymbol{\Pi}_e &= [0, 0, \Omega_e I_{zz}]^T, \boldsymbol{\gamma}_e = [0, 0, 1]^T, \boldsymbol{R}_e = [R_e \ 0 \ 0]^T, \\
\boldsymbol{P}_e &= [0 \ mR_e\Omega_e \ 0]^T, \boldsymbol{\Omega}_e = [0 \ 0 \ \Omega_e]^T.
\end{aligned} \tag{21}$$

Other relative equilibria can be converted into this equilibrium by changing the arrangement of the axes of the reference frame $S_b$.

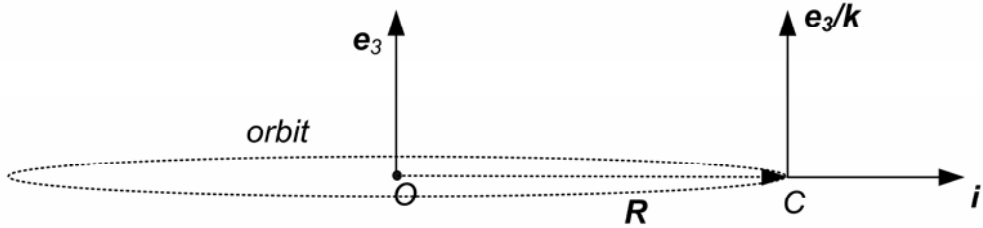

**Fig. 2.** One of the classical type of relative equilibria

## 3. Linear Stability of the Relative Equilibria

In this section, we will investigate the linear stability of the relative equilibria through the linear system matrix at the relative equilibria using the methods provided by the geometric mechanics (Beck and Hall 1998, Hall 2001).

**3.1 Conditions of linear stability**

The linear stability of the relative equilibrium $z_e$ depends on the eigenvalues of the linear system matrix of the system at the relative equilibrium. According to Beck and Hall (1998), the linear system matrix $D(z_e)$ of the non-canonical Hamiltonian system at the relative equilibrium $z_e$ can be calculated through the multiplication of the Poisson tensor and the Hessian of the variational Lagrangian without performing linearization as follows:

$$D(z_e) = B(z_e) \nabla^2 F(z_e). \tag{22}$$

Here the variational Lagrangian $F(z)$ is defined as

$$F(z) = H(z) - \sum_{i=1}^{2} \mu_i C_i(z). \tag{23}$$

According to Beck and Hall (1998), the relative equilibrium of the rigid body in the problem corresponds to the stationary point of the Hamiltonian constrained by the Casimir functions. The stationary points can be determined by the first variation condition of the variational Lagrangian $\nabla F(z_e) = 0$. By using the formulations of the Hamiltonian and Casimir functions, the equilibrium conditions are obtained as:

$$\begin{aligned}
&I^{-1} \Pi_e - \mu_2 \gamma_e = 0, \\
&\frac{3GM_1 \varepsilon m}{R_e^3} (\gamma_e \cdot \bar{R}_e) \bar{R}_e - \mu_1 \gamma_e - \mu_2 (\Pi_e + \hat{R}_e P_e) = 0, \\
&-\mu_2 \hat{P}_e \gamma_e + \left. \frac{\partial V}{\partial R} \right|_e = 0, \\
&-\mu_2 \hat{\gamma}_e R_e + \frac{P_e}{m} = 0.
\end{aligned} \tag{24}$$

As we expected, the relative equilibrium in Eq. (21) obtained based on the equations of motion is a solution of the equilibrium conditions Eq. (24), with the parameters $\mu_1$ and $\mu_2$ given by

$$\mu_1 = -\Omega_e^2 (I_{zz} + mR_e^2), \quad \mu_2 = \Omega_e. \tag{25}$$

By using the formulation of the second-order gravitational potential Eq. (8), the Hessian of the variational Lagrangian $\nabla^2 F(z)$ is calculated as:

$$\nabla^2 F(z) = \begin{bmatrix} \mathbf{I}^{-1} & -\mu_2 \mathbf{I}_{3\times 3} & \mathbf{0} & \mathbf{0} \\ -\mu_2 \mathbf{I}_{3\times 3} & \dfrac{3GM_1\varepsilon m}{R^5}\mathbf{R}\mathbf{R}^T - \mu_1 \mathbf{I}_{3\times 3} & \left(\dfrac{\partial^2 V}{\partial \gamma \partial \mathbf{R}}\right)^T + \mu_2 \hat{\mathbf{P}} & -\mu_2 \hat{\mathbf{R}} \\ \mathbf{0} & \dfrac{\partial^2 V}{\partial \gamma \partial \mathbf{R}} - \mu_2 \hat{\mathbf{P}} & \dfrac{\partial^2 V}{\partial \mathbf{R}^2} & \mu_2 \hat{\gamma} \\ \mathbf{0} & \mu_2 \hat{\mathbf{R}} & -\mu_2 \hat{\gamma} & \dfrac{1}{m}\mathbf{I}_{3\times 3} \end{bmatrix}. \quad (26)$$

The second-order partial derivates of the gravitational potential in Eq. (26) are obtained as follows:

$$\frac{\partial^2 V}{\partial \gamma \partial \mathbf{R}} = \frac{3GM_1\varepsilon m}{R^4}\left[(\gamma \cdot \overline{\mathbf{R}})\mathbf{I}_{3\times 3} + \gamma \overline{\mathbf{R}}^T - 5(\gamma \cdot \overline{\mathbf{R}})\overline{\mathbf{R}}\overline{\mathbf{R}}^T\right], \quad (27)$$

$$\begin{aligned}\frac{\partial^2 V}{\partial \mathbf{R}^2} &= \frac{GM_1 m}{R^3}\left(\mathbf{I}_{3\times 3} - 3\overline{\mathbf{R}}\overline{\mathbf{R}}^T\right) \\ &+ \frac{3GM_1}{2R^5}\left\{5\overline{\mathbf{R}}^T \mathbf{I}\overline{\mathbf{R}} - tr(\mathbf{I}) - \varepsilon m\left(1 - 5(\gamma \cdot \overline{\mathbf{R}})^2\right)\right\}\left\{7\overline{\mathbf{R}}\overline{\mathbf{R}}^T - \mathbf{I}_{3\times 3}\right\} \\ &+ \frac{3GM_1}{R^5}\left\{\left[tr(\mathbf{I}) + \varepsilon m\right]\overline{\mathbf{R}}\overline{\mathbf{R}}^T + \mathbf{I} + \varepsilon m \gamma \gamma^T\right\} \\ &+ \frac{15GM_1}{R^5}\left\{-\mathbf{I}\overline{\mathbf{R}}\overline{\mathbf{R}}^T - \overline{\mathbf{R}}\overline{\mathbf{R}}^T \mathbf{I} - \varepsilon m(\gamma \cdot \overline{\mathbf{R}})(\gamma \overline{\mathbf{R}}^T + \overline{\mathbf{R}}\gamma^T)\right\}.\end{aligned} \quad (28)$$

As described by Eqs. (17), (21) and (25), at the relative equilibrium $z_e$, we have $\boldsymbol{\Pi}_e = [0, 0, \Omega_e I_{zz}]^T$, $\gamma_e = [0, 0, 1]^T$, $\mathbf{R}_e = [R_e, 0, 0]^T$, $\overline{\mathbf{R}}_e = [1, 0, 0]^T$, $\mathbf{P}_e = [0, mR_e\Omega_e, 0]^T$, $\boldsymbol{\Omega}_e = [0, 0, \Omega_e]^T$, $\mu_1 = -\Omega_e^2(I_{zz} + mR_e^2)$ and $\mu_2 = \Omega_e$. Then the Hessian of the variational Lagrangian $\nabla^2 F(z_e)$ at the relative equilibrium $z_e$ can be obtained as:

$$\nabla^2 F(z_e) = \begin{bmatrix} \mathbf{I}^{-1} & -\mu_2 \mathbf{I}_{3\times 3} & \mathbf{0} & \mathbf{0} \\ -\mu_2 \mathbf{I}_{3\times 3} & \dfrac{3GM_1\varepsilon m}{R_e^5}\mathbf{R}_e\mathbf{R}_e^T - \mu_1 \mathbf{I}_{3\times 3} & \left(\dfrac{\partial^2 V}{\partial \gamma \partial \mathbf{R}}\right)^T\bigg|_e + \mu_2 \hat{\mathbf{P}}_e & -\mu_2 \hat{\mathbf{R}}_e \\ \mathbf{0} & \dfrac{\partial^2 V}{\partial \gamma \partial \mathbf{R}}\bigg|_e - \mu_2 \hat{\mathbf{P}}_e & \dfrac{\partial^2 V}{\partial \mathbf{R}^2}\bigg|_e & \mu_2 \hat{\gamma}_e \\ \mathbf{0} & \mu_2 \hat{\mathbf{R}}_e & -\mu_2 \hat{\gamma}_e & \dfrac{1}{m}\mathbf{I}_{3\times 3} \end{bmatrix}. \quad (29)$$

The second-order partial derivates of the gravitational potential in Eq. (29) at the relative equilibrium $z_e$ are obtained through Eqs. (27)-(28) as follows:

$$\left.\frac{\partial^2 V}{\partial \gamma \partial \mathbf{R}}\right|_e = \frac{3GM_1 \varepsilon m}{R_e^4} \gamma_e \alpha_e^T, \tag{30}$$

$$\left.\frac{\partial^2 V}{\partial \mathbf{R}^2}\right|_e = \frac{GM_1 m}{R_e^3}\left(\mathbf{I}_{3\times 3} - 3\alpha_e \alpha_e^T\right) + \frac{3GM_1}{2R_e^5}\left\{\begin{array}{l}\left[15I_{xx} - 5tr(\mathbf{I}) - 5\varepsilon m\right]\alpha_e \alpha_e^T + 2\varepsilon m \gamma_e \gamma_e^T \\ +2\mathbf{I} - \left[5I_{xx} - tr(\mathbf{I}) - \varepsilon m\right]\mathbf{I}_{3\times 3}\end{array}\right\}, \tag{31}$$

where $\alpha_e$ is defined as $\alpha_e = \begin{bmatrix} 1 & 0 & 0 \end{bmatrix}^T$.

The Poisson tensor $\mathbf{B}(z_e)$ at the relative equilibrium $z_e$ can be obtained as:

$$\mathbf{B}(z_e) = \begin{bmatrix} \Omega_e I_{zz} \hat{\gamma}_e & \hat{\gamma}_e & R_e \hat{\alpha}_e & mR_e \Omega_e \hat{\beta}_e \\ \hat{\gamma}_e & 0 & 0 & 0 \\ R_e \hat{\alpha}_e & 0 & 0 & \mathbf{E} \\ mR_e \Omega_e \hat{\beta}_e & 0 & -\mathbf{E} & 0 \end{bmatrix}, \tag{32}$$

where $\beta_e$ is defined as $\beta_e = \begin{bmatrix} 0, 1, 0 \end{bmatrix}^T$.

In Eqs. (29)-(32), we have

$$\hat{\alpha}_e = \begin{bmatrix} 0 & 0 & 0 \\ 0 & 0 & -1 \\ 0 & 1 & 0 \end{bmatrix}, \quad \hat{\beta}_e = \begin{bmatrix} 0 & 0 & 1 \\ 0 & 0 & 0 \\ -1 & 0 & 0 \end{bmatrix}, \quad \hat{\gamma}_e = \begin{bmatrix} 0 & -1 & 0 \\ 1 & 0 & 0 \\ 0 & 0 & 0 \end{bmatrix}, \quad \alpha_e \alpha_e^T = \begin{bmatrix} 1 & 0 & 0 \\ 0 & 0 & 0 \\ 0 & 0 & 0 \end{bmatrix}, \tag{33}$$

$$\gamma_e \gamma_e^T = \begin{bmatrix} 0 & 0 & 0 \\ 0 & 0 & 0 \\ 0 & 0 & 1 \end{bmatrix}, \quad \gamma_e \alpha_e^T = \begin{bmatrix} 0 & 0 & 0 \\ 0 & 0 & 0 \\ 1 & 0 & 0 \end{bmatrix}. \tag{34}$$

Then the linear system matrix $\mathbf{D}(z_e)$ of the non-canonical Hamiltonian system can be calculated through Eqs. (22), (29) and (32). Through some rearrangement and simplification, the linear system matrix $\mathbf{D}(z_e)$ can be written as follows:

$$\mathbf{D}(z_e) = \begin{bmatrix} \Omega_e I_{zz} \hat{\gamma}_e \mathbf{I}^{-1} \\ -\Omega_e \hat{\gamma}_e & 0 & \left\{\frac{GM_1 m}{R_e^2} - mR_e \Omega_e^2 - \frac{3GM_1}{2R_e^4}\left[5I_{xx} - tr(\mathbf{I}) - \varepsilon m\right]\right\}\hat{\alpha}_e & 0 \\ & & +\frac{3GM_1}{R_e^4}\hat{\alpha}_e \mathbf{I} \\ \hat{\gamma}_e \mathbf{I}^{-1} & -\Omega_e \hat{\gamma}_e & 0 & 0 \\ R_e \hat{\alpha}_e \mathbf{I}^{-1} & 0 & -\Omega_e \hat{\gamma}_e & \frac{1}{m}\mathbf{I}_{3\times 3} \\ mR_e \Omega_e \hat{\beta}_e \mathbf{I}^{-1} & -\left.\frac{\partial^2 V}{\partial \gamma \partial \mathbf{R}}\right|_e & -\left.\frac{\partial^2 V}{\partial \mathbf{R}^2}\right|_e & -\Omega_e \hat{\gamma}_e \end{bmatrix}. \tag{35}$$

As stated above, the linear stability of the relative equilibrium $z_e$ depends on the eigenvalues of the linear system matrix of the system $D(z_e)$. The characteristic polynomial of the linear system matrix $D(z_e)$ can be calculated by

$$P(s) = \det\left[ s\mathbf{I}_{12\times 12} - D(z_e) \right].  \tag{36}$$

The eigenvalues of the linear system matrix $D(z_e)$ are roots of the characteristic equation of the linearized system, which is given by

$$\det\left[ s\mathbf{I}_{12\times 12} - D(z_e) \right] = 0.  \tag{37}$$

Through Eqs. (35) and (37), with the help of *Matlab* and *Maple*, the characteristic equation can be obtained with the following form:

$$s^2 (m^2 I_{zz} s^4 + A_2 s^2 + A_0)(m I_{xx} I_{yy} s^6 + B_4 s^4 + B_2 s^2 + B_0) = 0,  \tag{38}$$

where the coefficients $A_2$, $A_0$, $B_4$, $B_2$ and $B_0$ are functions of the parameters of the system: $GM_1$, $\Omega_e$, $R_e$, $\varepsilon$, $m$, $I_{xx}$, $I_{yy}$ and $I_{zz}$. The explicit formulations of the coefficients are given in the Appendix.

According to Beck and Hall (1998), the non-canonical Hamiltonian systems have special properties with regard to both the form of the characteristic polynomial and the eigenvalues of the linear system matrix $D(z_e)$:

**Property 1.** *There are only even terms in the characteristic polynomial of the linear system matrix, and the eigenvalues are symmetrical with respect to both the real and imaginary axes.*

**Property 2.** *A zero eigenvalue exists for each linearly independent Casimir function.*

**Property 3.** *An additional pair of zero eigenvalues exists for each first integral, which is associated with a symmetry of the Hamiltonian by Noether's theorem.*

Notice that in our problem, there are two linearly independent Casimir functions, and the two zero eigenvalues correspond to the two Casimir functions $C_1(z)$ and $C_2(z)$. The remaining ten eigenvalues correspond to the motion constrained by the Casimir functions on the ten-dimensional invariant manifold $\Sigma$. We have carried out a Poisson reduction by means of the symmetry of the Hamiltonian, and expressed the dynamics on the reduced phase space. The additional pair of zero eigenvalues according to **Property 3** has been eliminated by the reduction process. Therefore, our results in Eq. (38) are consistent with these three properties stated above.

According to the characteristic equation in Eq. (38), the ten-dimensional linear system on the invariant manifold $\Sigma$ decouples into two entirely independent four- and six-dimensional subsystems under the second-order gravitational potential. It is worth our special attention that this is not the decoupling between the freedoms of the rotational motion and the orbital motion of the rigid body, since the orbit-rotation coupling is considered in our study. Actually, the four-dimensional subsystem and $s^2$ are the three freedoms of the orbital and rotational motions within the equatorial plane of the body *P*, and the other three freedoms, i.e. orbital and rotational motions outside the equatorial plane of the body *P*, constitute the six-dimensional subsystem.

The linear stability of the relative equilibria implies that there are no roots of the characteristic equation with positive real parts. According to **Property 1**, the linear stability requires all the roots to be purely imaginary, that is $s^2$ is real and negative. Therefore, in this case of a conservative system, we can only get the necessary conditions of the stability through the linear stability of the relative equilibria.

According to the theory of the roots of the second and third degree polynomial equation, that the $s^2$ in Eq. (38) is real and negative is equivalent to

$$\left(\frac{A_2}{m^2 I_{zz}}\right)^2 - \frac{4A_0}{m^2 I_{zz}} \geq 0, \ A_2 > 0, \ A_0 > 0; \tag{39}$$

$$\frac{1}{27}\left(-\frac{1}{3}\left(\frac{B_4}{mI_{xx}I_{yy}}\right)^2 + \frac{B_2}{mI_{xx}I_{yy}}\right)^3 + \frac{1}{4}\left(\frac{2}{27}\left(\frac{B_4}{mI_{xx}I_{yy}}\right)^3 - \frac{B_4 B_2}{3m^2 I_{xx}^2 I_{yy}^2} + \frac{B_0}{mI_{xx}I_{yy}}\right)^2 \leq 0, \tag{40}$$

$$B_4 > 0, \ B_2 > 0, \ B_0 > 0.$$

We have given the conditions of linear stability of the relative equilibria in Eqs. (39) and (40). Given a set of the parameters of the system, we can determine whether the relative equilibria are linear stability by using the stability criterion given above.

**3.2 Case studies**

However, the expressions of coefficients $A_2$, $A_0$, $B_4$, $B_2$ and $B_0$ in terms of the parameters of the system are tedious, since there are large amount of parameters in the system and the considered problem is a high-dimensional system. It is difficult to get general conditions of linear stability through Eqs. (39) and (40) in terms of the parameters of the system, i.e. $GM_1$, $\Omega_e$, $R_e$, $\varepsilon$, $m$, $I_{xx}$, $I_{yy}$ and $I_{zz}$.

We will consider an example planet $P$, which has the same mass and equatorial radius as the Earth, but has a different zonal harmonic $J_2$. That is $GM_1 = 3.986005 \times 10^{14} \text{m}^3/\text{s}^2$ and $a_E = 6.37814 \times 10^6 \text{m}$. Five different values of the zonal harmonic $J_2$ are considered

$$J_2 = 0.5, 0.2, 0, -0.18, -0.2. \tag{41}$$

The orbital angular velocity $\Omega_e$ is assumed to be equal to $1.163553 \times 10^{-3} \text{s}^{-1}$ with the orbital period equal to 1.5 hours.

With the parameters of the system given above, the stability criterion in Eqs. (39) and (40) can be determined by three mass distribution parameters of the rigid body: $I_{xx}/m$, $\sigma_x$ and $\sigma_y$, where $\sigma_x$ and $\sigma_y$ are defined as

$$\sigma_x = \left(\frac{I_{zz}-I_{yy}}{I_{xx}}\right), \quad \sigma_y = \left(\frac{I_{zz}-I_{xx}}{I_{yy}}\right). \tag{42}$$

The ratio $I_{xx}/m$ describes the characteristic dimension of the rigid body; the ratios $\sigma_x$ and $\sigma_y$ describe the shape of the rigid body to the second order. Three different values of the parameter $I_{xx}/m$ are considered as follows:

$$\frac{I_{xx}}{m} = 5\times 10^3, 5\times 10^7, 5\times 10^{11}, \tag{43}$$

which correspond to a rigid body with the characteristic dimension of order of 100m, 10km and 1000km respectively.

In the case of each value of $I_{xx}/m$, the parameters $\sigma_x$ and $\sigma_y$ are considered in the following range

$$-1 \leq \sigma_x \leq 1, -1 \leq \sigma_y \leq 1, \tag{44}$$

which have covered all the possible mass distributions of the rigid body.

Given the mass distribution parameters of the rigid body, the orbital radius $R_e$ at the relative equilibrium can be calculated by Eq. (17). Then the stability criterion in Eqs. (39) and (40) can be calculated with all the parameters of the system known.

The linear stability criterion in Eqs. (39) and (40) is calculated for a rigid body within the range of the parameters Eqs. (43) and (44) in the cases of different values of the zonal harmonic $J_2$. The points, which correspond to the mass distribution parameters guaranteeing linear stability, are plotted on the $\sigma_y - \sigma_x$ plane in the 15 cases of different values of $I_{xx}/m$ and $J_2$ in Figs. (3)-(17) respectively.

In our problem, the gravitational potential in Eq. (8) is truncated on the second order. According to the conclusions in Wang and Xu (2013b), only the central component of the gravity field of the planet *P* is considered in the gravity gradient torque, with the zonal harmonic $J_2$ neglected. That is to say, the attitude motion of the rigid body in our problem, in the point view of the traditional attitude dynamics with the orbit-rotation coupling neglected, is actually the attitude dynamics on a circular orbit in a central gravity field. To make comparisons with the traditional attitude dynamics, we also plot the classical linear attitude stability region of a rigid body on a circular orbit in a central gravity field in Figs. (3)-(17), which is given by:

$$\begin{aligned} \sigma_y - \sigma_x &> 0, \\ 1 + 3\sigma_y + \sigma_x \sigma_y &> 4\sqrt{\sigma_x \sigma_y}, \\ \sigma_x \sigma_y &> 0. \end{aligned} \quad (45)$$

The classical linear attitude stability region given by Eq. (45) is consisted of the Lagrange region I and the DeBra-Delp region II (Hughes 1986). The Lagrange region is the isosceles right triangle region in the first quadrant of the $\sigma_y - \sigma_x$ plane below the straight line $\sigma_y - \sigma_x = 0$, and DeBra-Delp region is a small region in the third quadrant below the straight line $\sigma_y - \sigma_x = 0$.

Notice that at the relative equilibrium in our paper, the orientations of the principal axes of the rigid body are different from those at the equilibrium attitude in Hughes (1986), and then the definitions of the parameters $\sigma_y$ and $\sigma_x$ in our paper are different form those in Hughes (1986) to make sure that the linear attitude stability region is the same as in Hughes (1986).

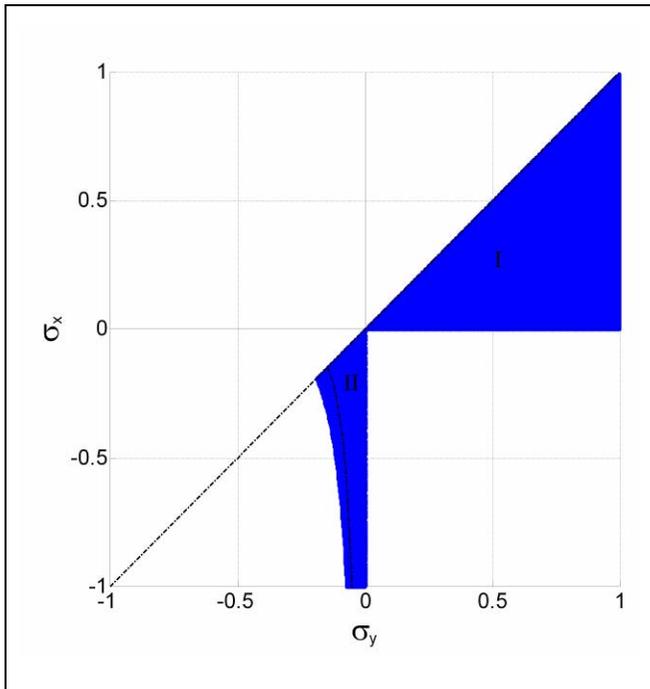 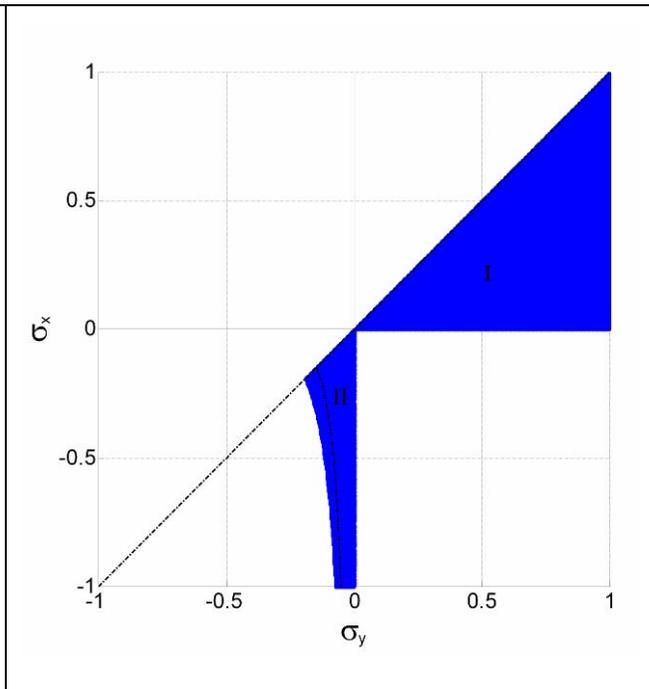 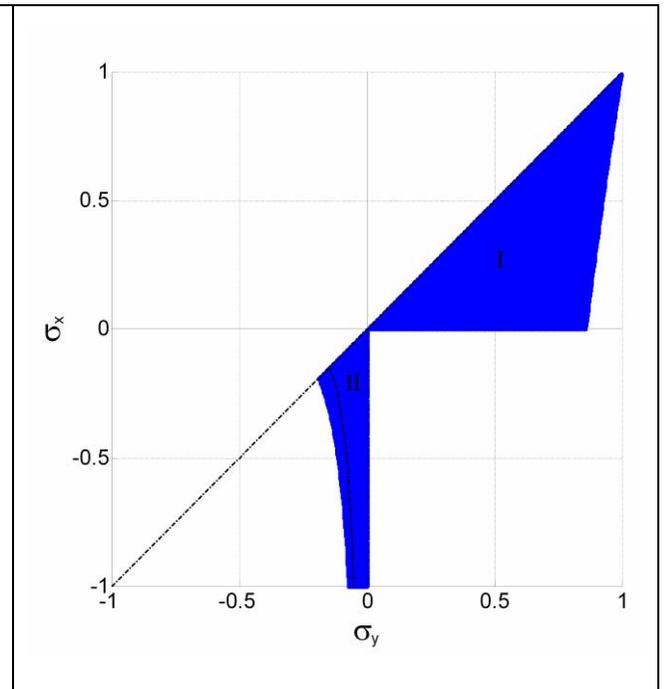

| **Fig. 3.** Linear stability region on $\sigma_y - \sigma_x$ plane in the case of $J_2 = 0.5$ and $I_{xx}/m = 5 \times 10^3$ | **Fig. 4.** Linear stability region on $\sigma_y - \sigma_x$ plane in the case of $J_2 = 0.5$ and $I_{xx}/m = 5 \times 10^7$ | **Fig. 5.** Linear stability region on $\sigma_y - \sigma_x$ plane in the case of $J_2 = 0.5$ and $I_{xx}/m = 5 \times 10^{11}$ |
|---|---|---|

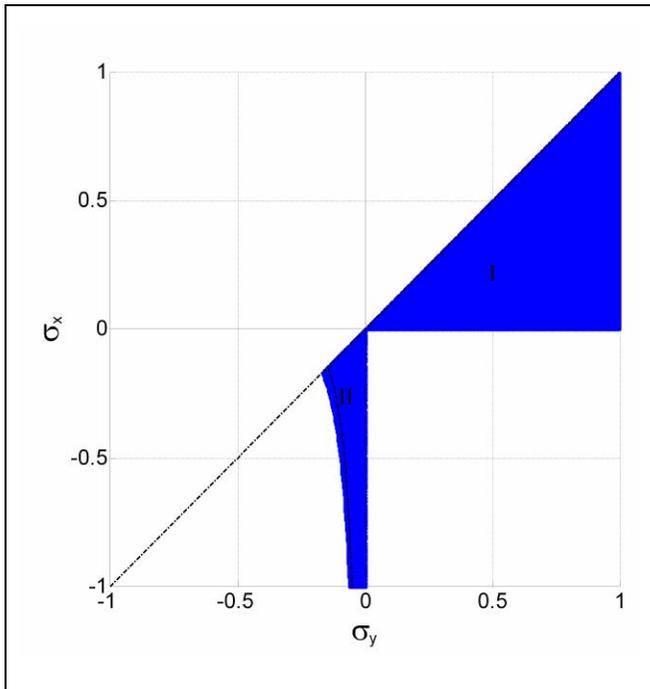 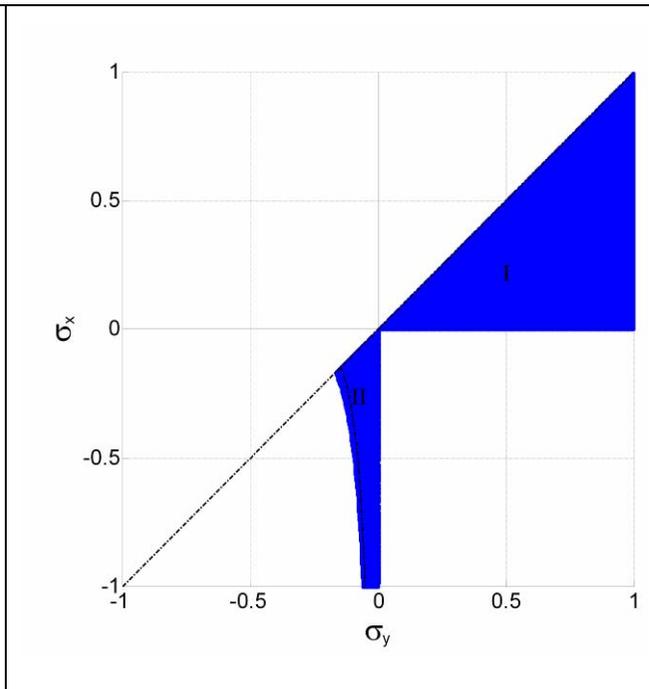 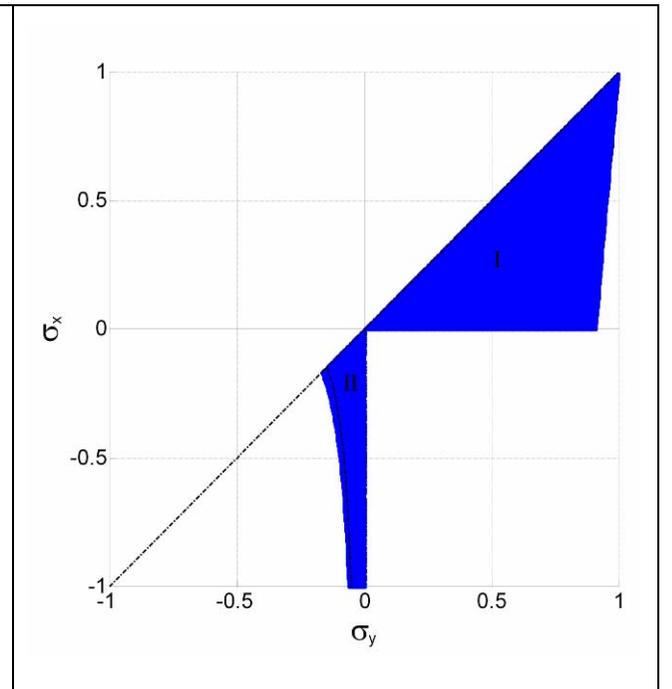

| **Fig. 6.** Linear stability region on $\sigma_y - \sigma_x$ plane in the case of $J_2 = 0.2$ and $I_{xx}/m = 5 \times 10^3$ | **Fig. 7.** Linear stability region on $\sigma_y - \sigma_x$ plane in the case of $J_2 = 0.2$ and $I_{xx}/m = 5 \times 10^7$ | **Fig. 8.** Linear stability region on $\sigma_y - \sigma_x$ plane in the case of $J_2 = 0.2$ and $I_{xx}/m = 5 \times 10^{11}$ |

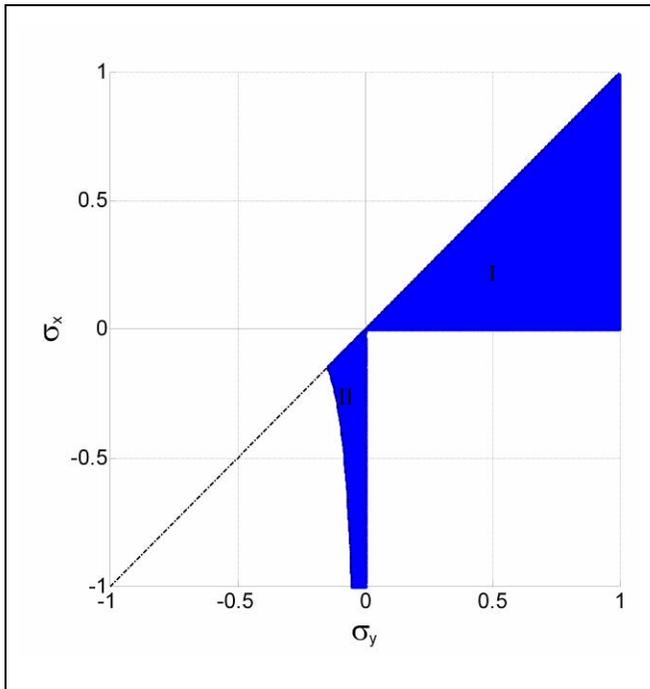
**Fig. 9.** Linear stability region on $\sigma_y - \sigma_x$ plane in the case of $J_2 = 0$ and $I_{xx}/m = 5 \times 10^3$

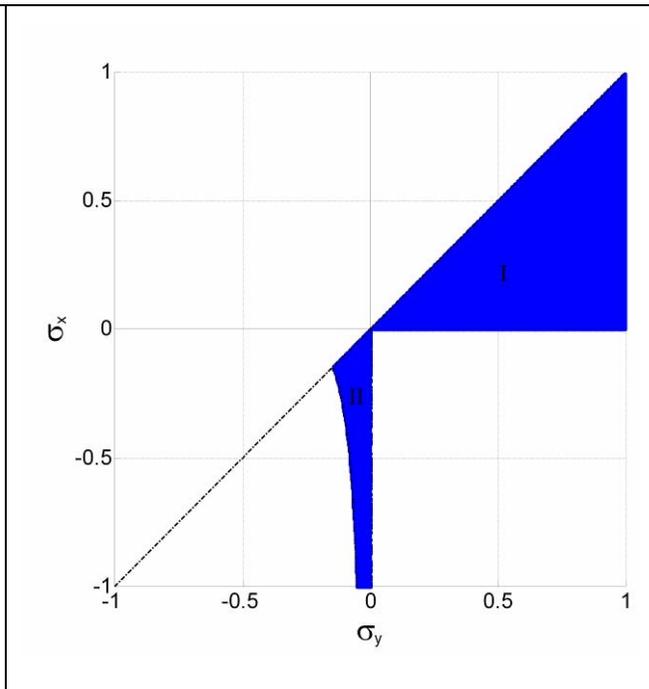
**Fig. 10.** Linear stability region on $\sigma_y - \sigma_x$ plane in the case of $J_2 = 0$ and $I_{xx}/m = 5 \times 10^7$

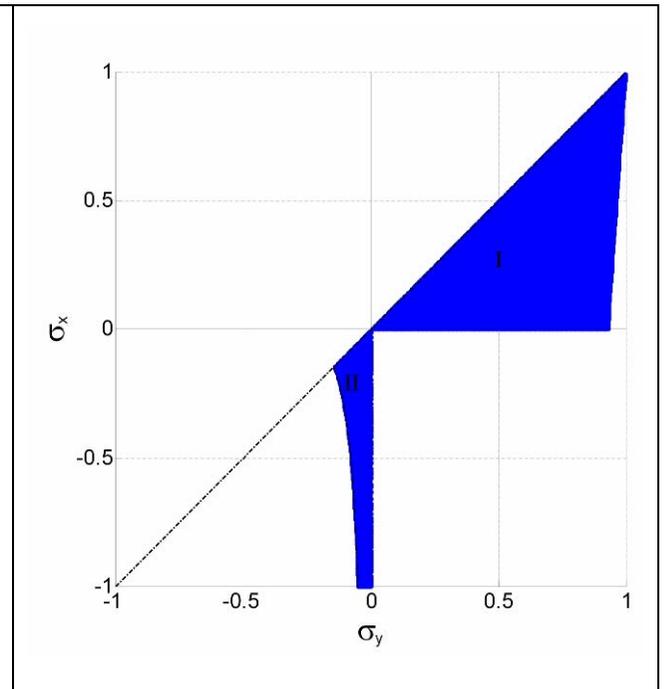
**Fig. 11.** Linear stability region on $\sigma_y - \sigma_x$ plane in the case of $J_2 = 0$ and $I_{xx}/m = 5 \times 10^{11}$

| 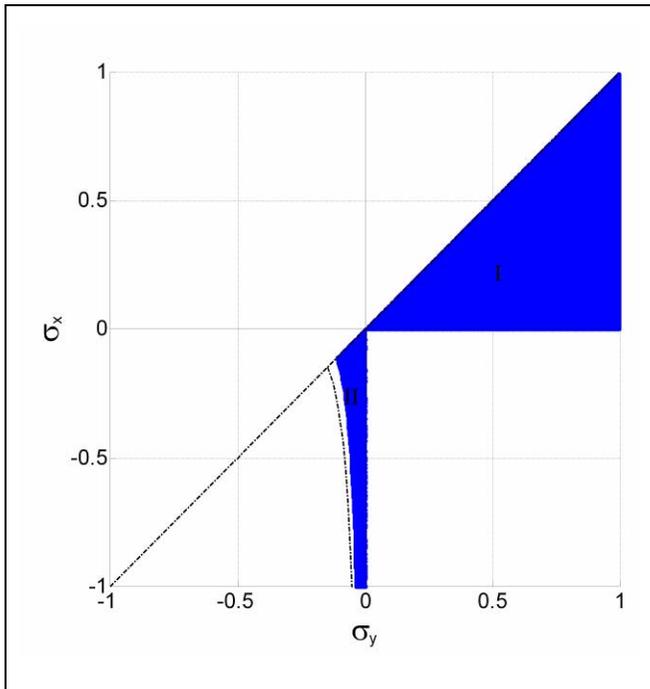 | 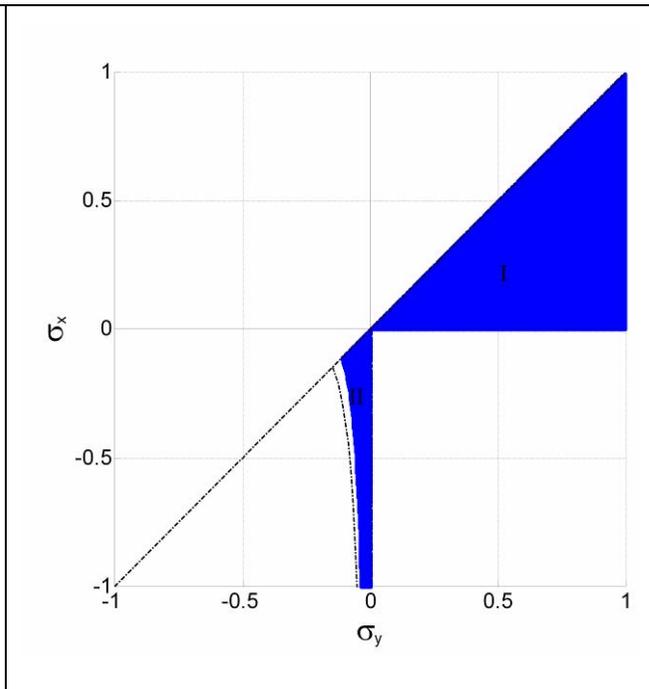 | 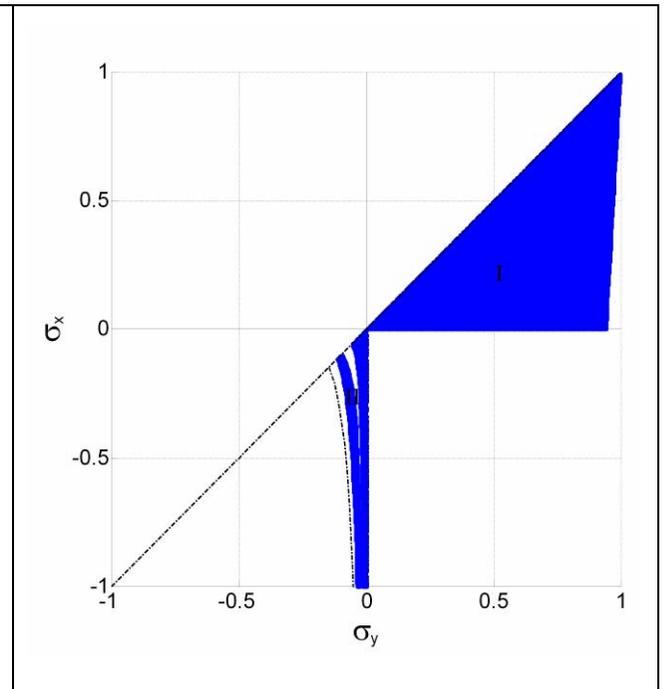 |
|---|---|---|
| **Fig. 12.** Linear stability region on $\sigma_y - \sigma_x$ plane in the case of $J_2 = -0.18$ and $I_{xx}/m = 5 \times 10^3$ | **Fig. 13.** Linear stability region on $\sigma_y - \sigma_x$ plane in the case of $J_2 = -0.18$ and $I_{xx}/m = 5 \times 10^7$ | **Fig. 14.** Linear stability region on $\sigma_y - \sigma_x$ plane in the case of $J_2 = -0.18$ and $I_{xx}/m = 5 \times 10^{11}$ |

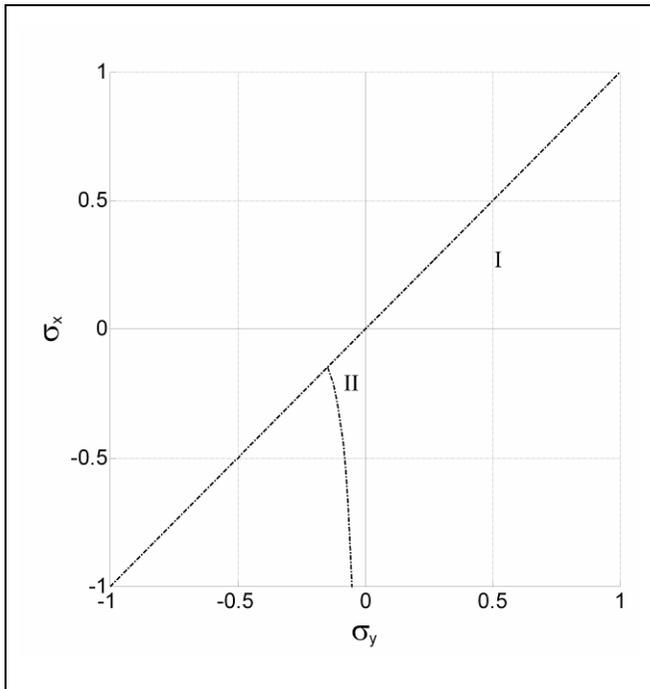 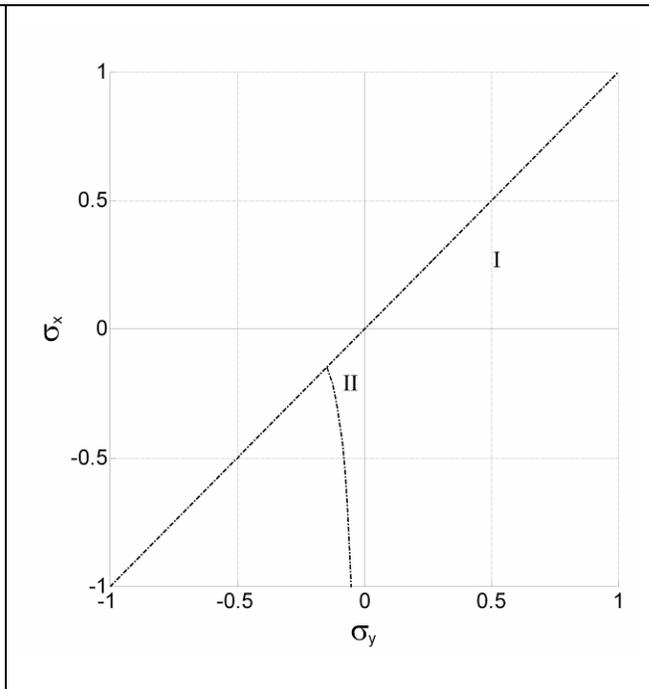 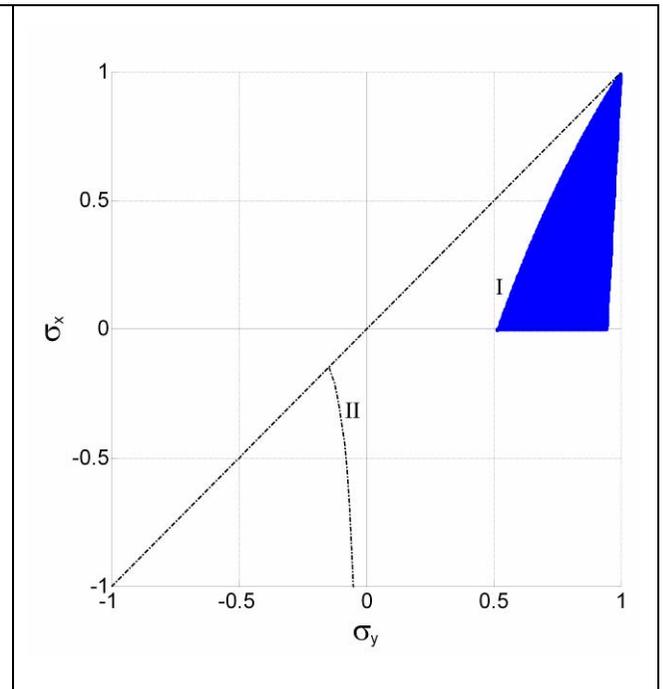

| **Fig. 15.** Linear stability region on $\sigma_y - \sigma_x$ plane in the case of $J_2 = -0.2$ and $I_{xx}/m = 5 \times 10^3$ | **Fig. 16.** Linear stability region on $\sigma_y - \sigma_x$ plane in the case of $J_2 = -0.2$ and $I_{xx}/m = 5 \times 10^7$ | **Fig. 17.** Linear stability region on $\sigma_y - \sigma_x$ plane in the case of $J_2 = -0.2$ and $I_{xx}/m = 5 \times 10^{11}$ |
|---|---|---|

**3.3 Some discussions on the linear stability**

From Figs. (3)-(17), we can easily achieve several conclusions as follows:

(a). Similar to the classical linear attitude stability region, which is consisted of the Lagrange region and the DeBra-Delp region, the linear stability region of the relative equilibrium of the rigid body in our problem is also consisted of two regions located in the first and third quadrant of the $\sigma_y - \sigma_x$ plane respectively, which are the analogues of the Lagrange region and the DeBra-Delp region respectively. This is consistent with the conclusion by Teixidó Román (2010) that for a rigid body in a central gravity field there is a linear stability region in the third quadrant of the $\sigma_y - \sigma_x$ plane, which is the analogue of the DeBra-Delp region.

However, when the planet $P$ is very elongated with $J_2 = -0.2$, for a small rigid body there is no linear stability region; only in the case of a very large rigid body with $I_{xx}/m = 5 \times 10^{11}$, there is a linear stability region that is the analogue of the Lagrange region located in the first quadrant of the $\sigma_y - \sigma_x$ plane.

(b). For a given value of the zonal harmonic $J_2$ (except $J_2 = -0.2$), when the characteristic dimension of the rigid body is small, the characteristic dimension of the rigid body have no influence on the linear stability region, as shown by the linear stability region in the cases of $I_{xx}/m = 5 \times 10^3$ and $I_{xx}/m = 5 \times 10^7$. In these cases, the linear stability region in the first quadrant of the $\sigma_y - \sigma_x$ plane, the analogue of the Lagrange region, is actually the Lagrange region.

When the characteristic dimension of the rigid body is large enough, such as $I_{xx}/m = 5 \times 10^{11}$, the linear stability region in the first quadrant of the $\sigma_y - \sigma_x$ plane, the analogue of the Lagrange region, is reduced by a triangle in the right part of the first quadrant of the $\sigma_y - \sigma_x$ plane, as shown by Figs. (5), (8), (11) and (14). In the case of $J_2 = -0.18$, also the linear stability region in the third quadrant of the $\sigma_y - \sigma_x$ plane, the analogue of the DeBra-Delp region, is reduced by the large characteristic dimension of the rigid body, as shown by Fig. (14).

(c). For a given value of the characteristic dimension of the rigid body, as the zonal harmonic $J_2$ increases from -0.18 to 0.5, the linear stability region in the third quadrant of the $\sigma_y - \sigma_x$ plane, the analogue of the DeBra-Delp region, expands in the direction of the boundary of the DeBra-Delp region, and cross the boundary of the DeBra-Delp region at $J_2 = 0$.

For a small value of the characteristic dimension of the rigid body, such as $I_{xx}/m = 5 \times 10^3$ and $I_{xx}/m = 5 \times 10^7$, as the zonal harmonic $J_2$ increases from -0.18 to 0.5, the linear stability region in the first quadrant of the $\sigma_y - \sigma_x$ plane, the analogue of the Lagrange region, keeps equal to the Lagrange region. Whereas for a large value of the characteristic dimension of the rigid body $I_{xx}/m = 5 \times 10^{11}$, as the zonal harmonic $J_2$ increases from -0.18 to 0.5, the linear stability region in the first quadrant of the $\sigma_y - \sigma_x$ plane, the analogue of the Lagrange region, shrinks by the influence of the zonal harmonic $J_2$.

## 4. Nonlinear Stability of the Relative Equilibria

In this section, we will investigate the nonlinear stability of the classical type of relative equilibria using the energy-Casimir method provided by the geometric mechanics adopted by Beck and Hall (1998), and Hall (2001).

**4.1 Conditions of nonlinear stability**

The energy-Casimir method, the generalization of Lagrange-Dirichlet criterion, is a powerful tool provided by the geometric mechanics for determining the nonlinear stability of the relative equilibria in a non-canonical Hamiltonian system (Marsden and Ratiu, 1999). According to the Lagrange-Dirichlet criterion in the canonical Hamiltonian system, the nonlinear stability of the equilibrium point is determined by the distributions of the eigenvalues of the Hessian matrix of the Hamiltonian. If all the eigenvalues of the Hessian matrix are positive or negative, that is the Hessian matrix of the Hamiltonian is positive- or negative-definite, then the equilibrium point is nonlinear stable. This follows from the conservation of energy and the fact that the level sets of the Hamiltonian near the equilibrium point are approximately ellipsoids.

However, the Hamiltonian system in our problem is non-canonical, and the phase flow of the system is constrained on the ten-dimensional invariant manifold or symplectic leaf $\Sigma$ by Casimir functions. Therefore, rather than considering general perturbations in the whole phase space as in the Lagrange-Dirichlet criterion in the canonical Hamiltonian system, we need to restrict the consideration to perturbations on $T\Sigma|_{z_e}$, the tangent space to the invariant manifold $\Sigma$ at the relative equilibrium $z_e$. $T\Sigma|_{z_e}$ is also the range space of Poisson tensor $B(z)$ at the relative equilibrium $z_e$, denoted by $\text{R}(B(z_e))$. This is the basic principle of the energy-Casimir method that the Hessian matrix needs to be considered restrictedly on the invariant manifold $\Sigma$ of the system. This restriction is constituted through the projected Hessian matrix of the variational Lagrangian $F(z)$ in Beck and Hall (1998).

According to the energy-Casimir method adopted by Beck and Hall (1998), the conditions of nonlinear stability of the relative equilibrium $z_e$ can be obtained through the distributions of the eigenvalues of the projected Hessian matrix of the variational Lagrangian $F(z)$. The projected Hessian matrix of the variational Lagrangian $F(z)$ has the same number of zero eigenvalues as the linearly independent Casimir functions, which are associated with the nullspace $\text{N}[B(z_e)]$, i.e. the complement space of $T\Sigma|_{z_e}$. The remaining eigenvalues of the projected Hessian matrix are associated with the tangent space to the invariant manifold $T\Sigma|_{z_e}$. If they are all positive, the relative equilibrium $z_e$ is a constrained minimum on the invariant manifold $\Sigma$ and therefore it is nonlinear stable.

According to Beck and Hall (1998), the projected Hessian matrix is given by $P(z_e)\nabla^2 F(z_e)P(z_e)$, where the projection operator is given by

$$P(z_e) = \mathbf{I}_{12\times12} - K(z_e)\left(K(z_e)^T K(z_e)\right)^{-1} K(z_e)^T. \quad (46)$$

As described by Eqs. (17), (21) and (25), at the relative equilibrium $z_e$, we have

$$K(z_e) = \text{N}[B(z_e)] = \begin{bmatrix} \mathbf{0} & \gamma_e \\ \gamma_e & \Pi_e + \hat{R}_e P_e \\ \mathbf{0} & \hat{P}_e \gamma_e \\ \mathbf{0} & \hat{\gamma}_e R_e \end{bmatrix}. \quad (47)$$

Using the Hessian of the variational Lagrangian $\nabla^2 F(z_e)$ given by Eq. (29) and the projection operator $P(z_e)$, we can calculate the projected Hessian matrix $P(z_e)\nabla^2 F(z_e)P(z_e)$.

As stated above, the nonlinear stability of the relative equilibrium $z_e$ depends on the eigenvalues of the projected Hessian matrix of the variational Lagrangian $F(z)$. The characteristic polynomial of the projected Hessian matrix $P(z_e)\nabla^2 F(z_e)P(z_e)$

can be calculated by

$$Q(s) = \det\left[ s\mathbf{I}_{12\times12} - \boldsymbol{P}(z_e)\nabla^2 F(z_e)\boldsymbol{P}(z_e) \right]. \qquad (48)$$

The eigenvalues of the projected Hessian matrix are roots of the characteristic equation, which is given by

$$\det\left[ s\mathbf{I}_{12\times12} - \boldsymbol{P}(z_e)\nabla^2 F(z_e)\boldsymbol{P}(z_e) \right] = 0. \qquad (49)$$

Through Eq. (49), with the help of *Matlab* and *Maple*, the characteristic equation can be obtained with the following form:

$$s^2(s^3 + C_2 s^2 + C_1 s + C_0)(s^3 + D_2 s^2 + D_1 s + D_0)(s^2 + E_1 s + E_0)(s^2 + F_1 s + F_0) = 0, \qquad (50)$$

where coefficients $C_2$, $C_1$, $C_0$, $D_2$, $D_1$, $D_0$, $E_1$, $E_0$, $F_1$ and $F_0$ are functions of the parameters of the system: $GM_1$, $\Omega_e$, $R_e$, $\varepsilon$, $m$, $I_{xx}$, $I_{yy}$ and $I_{zz}$. The explicit formulations of the coefficients are given in the Appendix.

In our problem there are two linearly independent Casimir functions, then as shown by Eq. (50), the projected Hessian matrix have two zero eigenvalues associated with the two-dimensional complement space of $T\Sigma|_{z_e}$. The remaining ten eigenvalues are associated with the ten-dimensional tangent space $T\Sigma|_{z_e}$ to the invariant manifold, and if they are all positive, then the relative equilibrium $z_e$ is a constrained minimum on the invariant manifold $\Sigma$, therefore it is nonlinear stable.

Since the projected Hessian matrix is symmetrical, the eigenvalues are guaranteed to be real by the coefficients of the polynomials in Eq. (50) intrinsically. Therefore, in the conditions of nonlinear stability of the relative equilibria, it is only needed to guarantee that the roots of the polynomial equations in Eq. (50) are positive.

According to the theory of roots of the polynomial equation, that the remaining ten eigenvalues in Eq. (50) are positive is equivalent to

$$\begin{array}{c} C_2 < 0, C_1 > 0, C_0 < 0, \\ D_2 < 0, D_1 > 0, D_0 < 0, \\ E_1 < 0, E_0 > 0, \\ F_1 < 0, F_0 > 0. \end{array} \quad (51)$$

We have given the conditions of the nonlinear stability of the relative equilibria in Eq. (51). Given the parameters of the system, we can determine whether the relative equilibria are nonlinear stability using the stability criterion in Eq. (51).

**4.2 Case studies**

As in the studies of the linear stability, here we also give case studies using numerical method. The parameters of the problem considered here are same as in the linear stability studies.

We calculate the nonlinear stability criterion in Eqs. (51) for a rigid body within the range of the parameters given by Eqs. (43) and (44) in the cases of five different values of the zonal harmonic $J_2$ given by Eq. (41). The points, which correspond to the mass distribution parameters guaranteeing the nonlinear stability, are plotted on the $\sigma_y - \sigma_x$ plane in the 15 cases of different values of $I_{xx}/m$ and $J_2$ in Figs. (18)-(32) respectively.

To make comparisons with the traditional attitude dynamics, we have also given the classical nonlinear attitude stability region of a rigid body on a circular orbit in a central gravity field in the Figs. (18)-(32), which is the Lagrange region, the isosceles right triangle region in the first quadrant of the $\sigma_y - \sigma_x$ plane below the straight line $\sigma_y - \sigma_x = 0$.

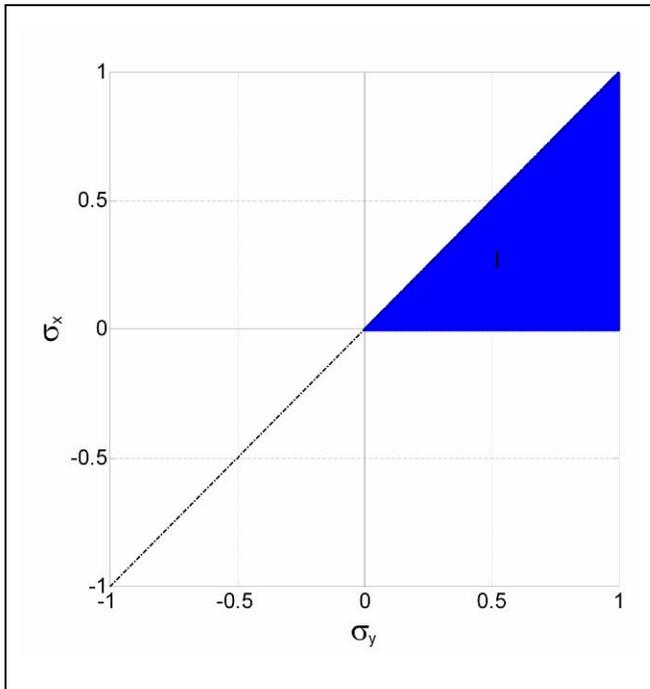 | 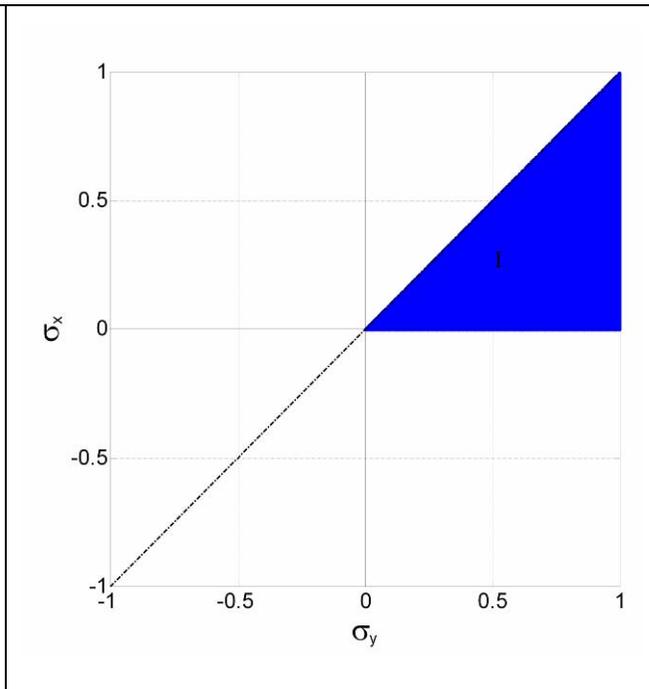 | 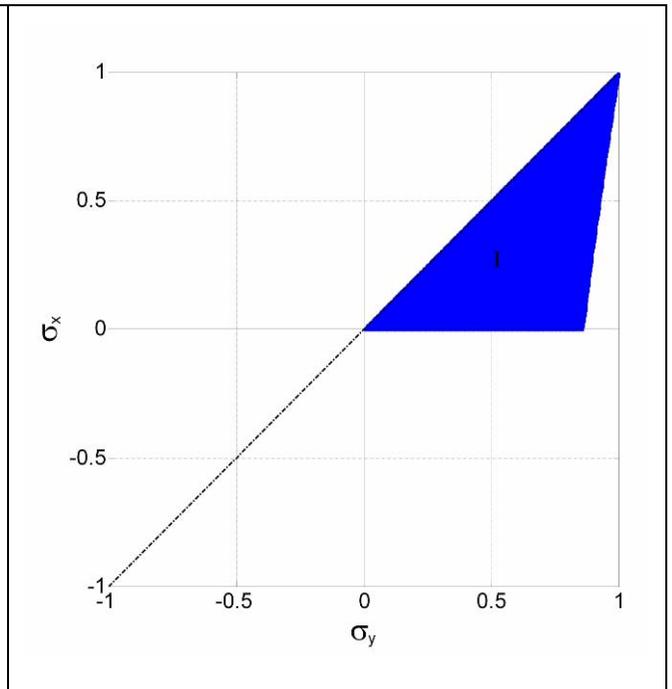

**Fig. 18.** Nonlinear stability region on $\sigma_y - \sigma_x$ plane in the case of $J_2 = 0.5$ and $I_{xx}/m = 5 \times 10^3$

**Fig. 19.** Nonlinear stability region on $\sigma_y - \sigma_x$ plane in the case of $J_2 = 0.5$ and $I_{xx}/m = 5 \times 10^7$

**Fig. 20.** Nonlinear stability region on $\sigma_y - \sigma_x$ plane in the case of $J_2 = 0.5$ and $I_{xx}/m = 5 \times 10^{11}$

| 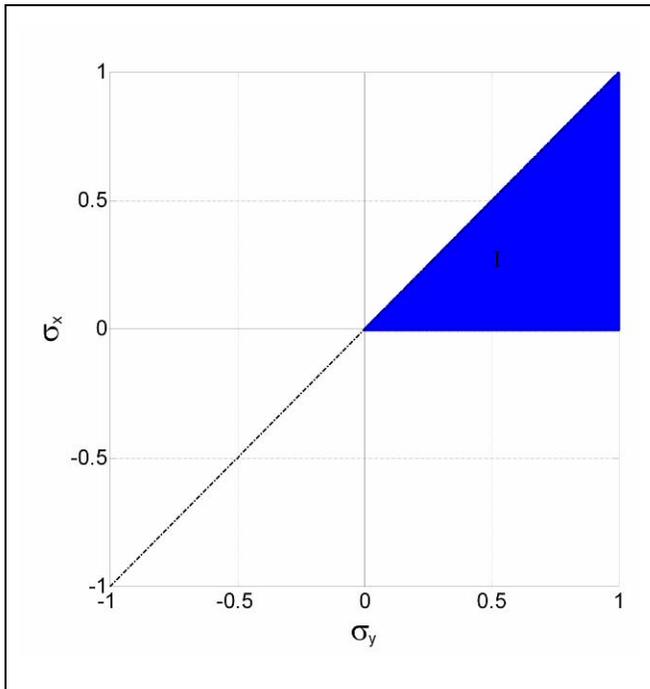 | 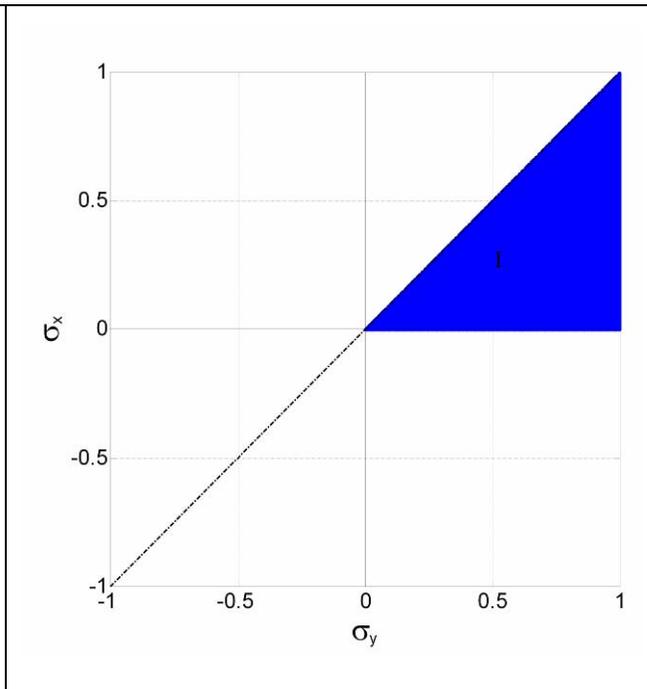 | 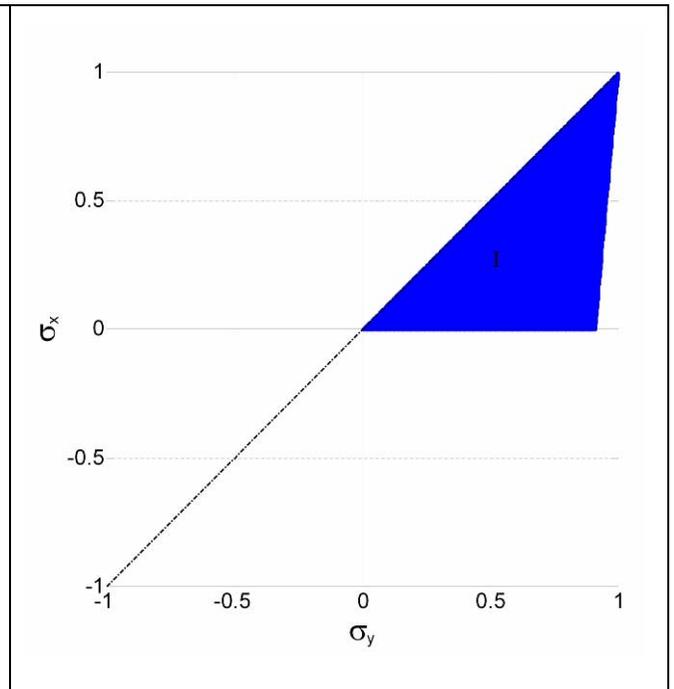 |
|---|---|---|
| **Fig. 21.** Nonlinear stability region on $\sigma_y - \sigma_x$ plane in the case of $J_2 = 0.2$ and $I_{xx}/m = 5 \times 10^3$ | **Fig. 22.** Nonlinear stability region on $\sigma_y - \sigma_x$ plane in the case of $J_2 = 0.2$ and $I_{xx}/m = 5 \times 10^7$ | **Fig. 23.** Nonlinear stability region on $\sigma_y - \sigma_x$ plane in the case of $J_2 = 0.2$ and $I_{xx}/m = 5 \times 10^{11}$ |

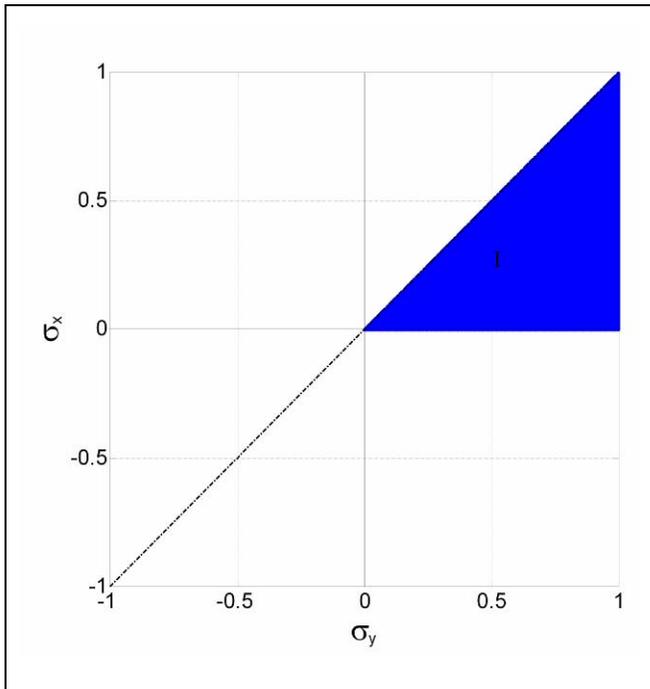 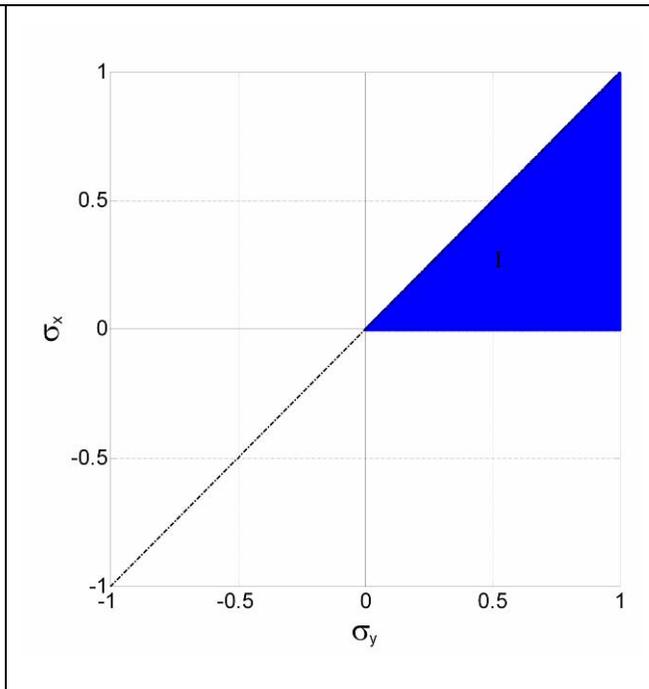 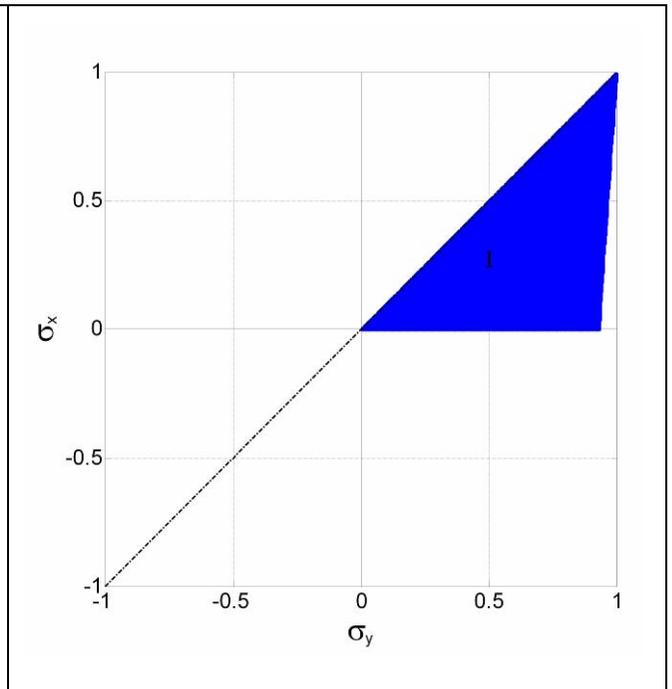

| **Fig. 24.** Nonlinear stability region on $\sigma_y - \sigma_x$ plane in the case of $J_2 = 0$ and $I_{xx}/m = 5 \times 10^3$ | **Fig. 25.** Nonlinear stability region on $\sigma_y - \sigma_x$ plane in the case of $J_2 = 0$ and $I_{xx}/m = 5 \times 10^7$ | **Fig. 26.** Nonlinear stability region on $\sigma_y - \sigma_x$ plane in the case of $J_2 = 0$ and $I_{xx}/m = 5 \times 10^{11}$ |

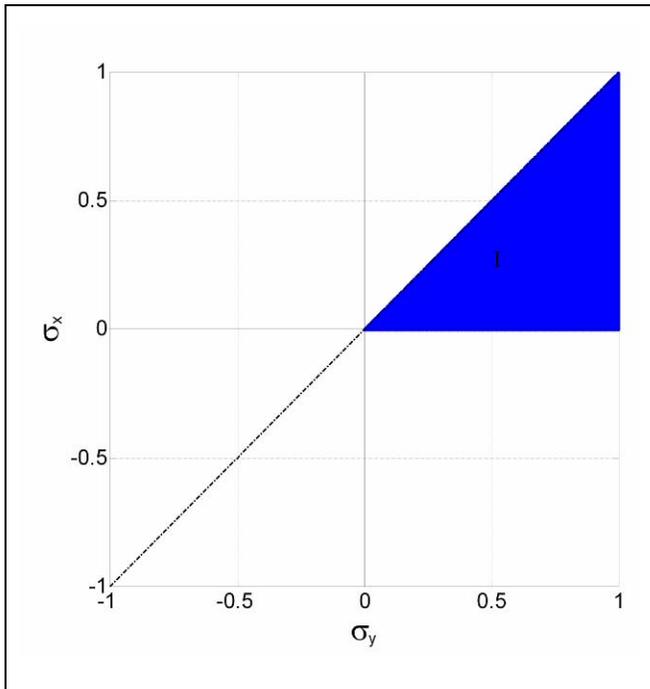 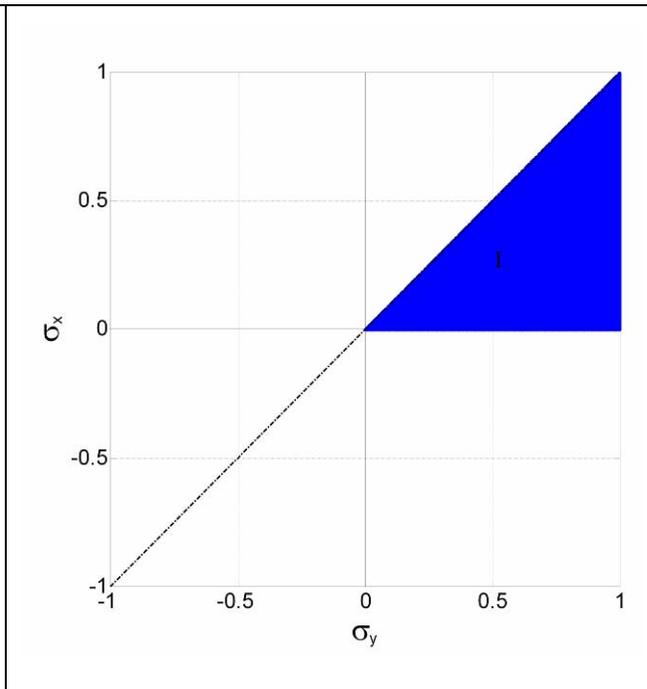 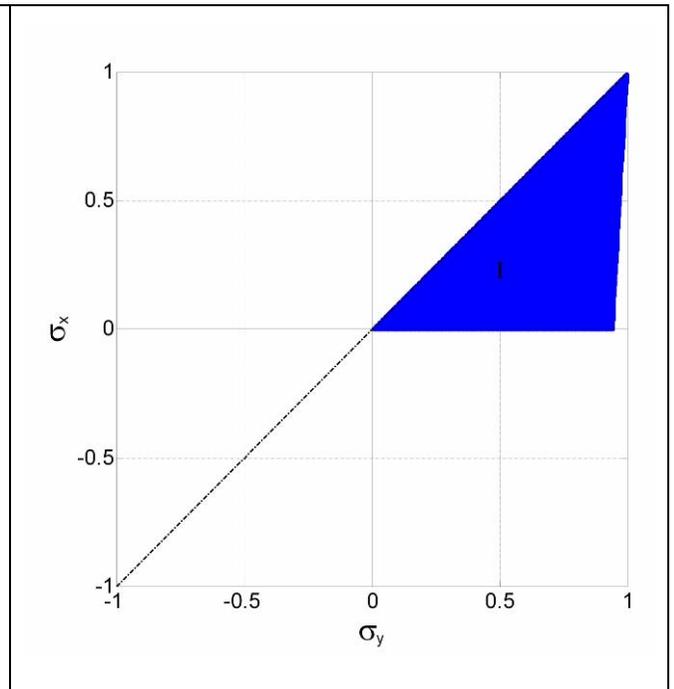

| **Fig. 27.** Nonlinear stability region on $\sigma_y - \sigma_x$ plane in the case of $J_2 = -0.18$ and $I_{xx}/m = 5 \times 10^3$ | **Fig. 28.** Nonlinear stability region on $\sigma_y - \sigma_x$ plane in the case of $J_2 = -0.18$ and $I_{xx}/m = 5 \times 10^7$ | **Fig. 29.** Nonlinear stability region on $\sigma_y - \sigma_x$ plane in the case of $J_2 = -0.18$ and $I_{xx}/m = 5 \times 10^{11}$ |

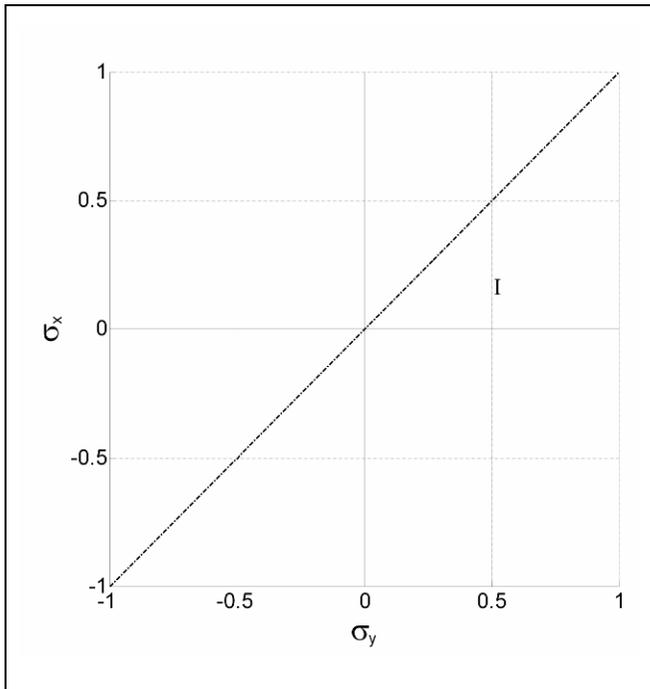 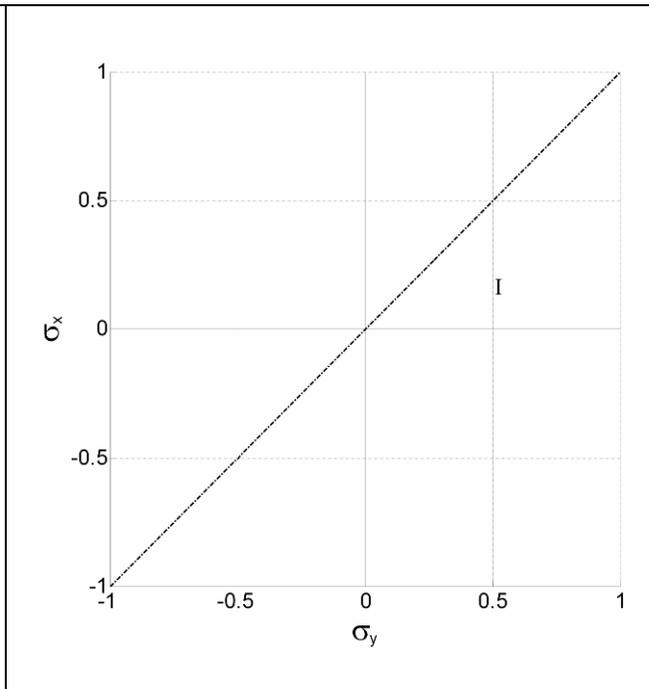 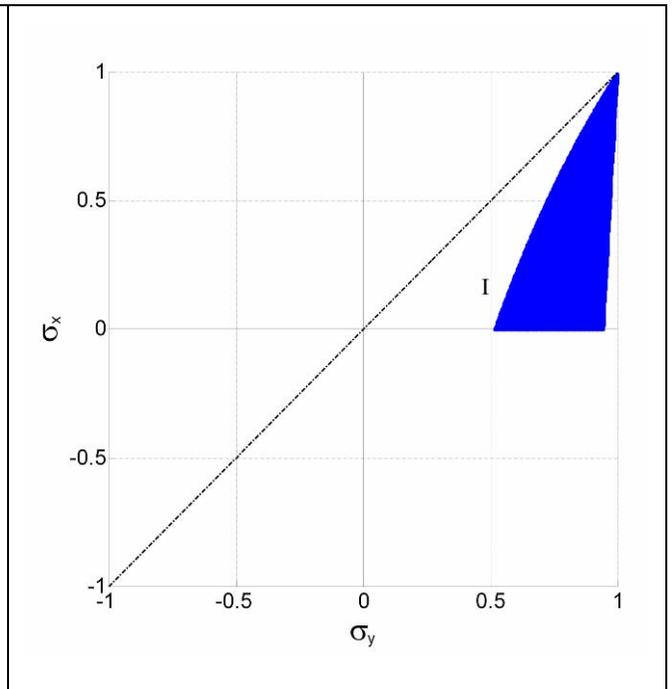

| **Fig. 30.** Nonlinear stability region on $\sigma_y - \sigma_x$ plane in the case of $J_2 = -0.2$ and $I_{xx}/m = 5 \times 10^3$ | **Fig. 31.** Nonlinear stability region on $\sigma_y - \sigma_x$ plane in the case of $J_2 = -0.2$ and $I_{xx}/m = 5 \times 10^7$ | **Fig. 32.** Nonlinear stability region on $\sigma_y - \sigma_x$ plane in the case of $J_2 = -0.2$ and $I_{xx}/m = 5 \times 10^{11}$ |

**4.3 Some discussions on the nonlinear stability**

From Figs. (18)-(32), we can easily achieve several conclusions as follows:

(a). In all the 15 cases of different values of $I_{xx}/m$ and $J_2$, the nonlinear stability region is the subset of the linear stability region in the first quadrant that is the analogue of the Lagrange region. This is similar to the classical attitude stability problem of a rigid body in a central gravity field, in which the nonlinear attitude stability region is also the subset of the linear attitude stability region in the first quadrant, i.e., the Lagrange region. This is consistent with the stability theory of the Hamiltonian system that the linear stability is the necessary condition of the stability whereas the nonlinear stability is the sufficient condition of the stability, and the sufficient stability condition should be a subset of the necessary stability condition.

When the planet P is very elongated with $J_2 = -0.2$, for a small rigid body there is no linear stability region and then there is no nonlinear stability region; only for a very large rigid body with $I_{xx}/m = 5 \times 10^{11}$, there is a linear stability region, which is also a nonlinear stability region, located in the first quadrant of the $\sigma_y - \sigma_x$ plane.

(b). For a given value of the zonal harmonic $J_2$ (except $J_2 = -0.2$), when the characteristic dimension of the rigid body is small, the characteristic dimension of the rigid body have no influence on the nonlinear stability region, as shown by the nonlinear stability region in the cases of $I_{xx}/m = 5 \times 10^3$ and $I_{xx}/m = 5 \times 10^7$. In these cases, the nonlinear stability region is actually the Lagrange region, which is consistent with conclusions by Wang et al. (1991) and Teixidó Román (2010) on the rigid body dynamics in a central gravity filed.

When the characteristic dimension of the rigid body is large enough, such as $I_{xx}/m = 5\times 10^{11}$, the nonlinear stability region, the Lagrange region, is reduced by a triangle in the right part of the first quadrant of the $\sigma_y - \sigma_x$ plane, as shown by Figs. (5), (8), (11) and (14). As the zonal harmonic $J_2$ increases from -0.18 to 0.5, the reduction of the Lagrange region expands and the nonlinear stability region shrinks. Notice that even in a central gravity field with $J_2 = 0$, the nonlinear stability region is not the Lagrange region anymore. This result has not been obtained in previous works, such as Wang et al. (1991) and Teixidó Román (2010).

(c). For a small characteristic dimension of the rigid body, such as $I_{xx}/m = 5\times 10^3$ and $I_{xx}/m = 5\times 10^7$, as the zonal harmonic $J_2$ increases from -0.18 to 0.5, the nonlinear stability region keeps equal to the Lagrange region. Whereas for a large value of the characteristic dimension of the rigid body $I_{xx}/m = 5\times 10^{11}$, as the zonal harmonic $J_2$ increases from -0.18 to 0.5, the nonlinear stability region shrinks by the influence of the zonal harmonic $J_2$.

## 5. Conclusions

For new high-precision applications in celestial mechanics and astrodynamics, we have generalized the classical $J_2$ problem to the motion of a rigid body in a $J_2$ gravity field. Based on our previous results on the relative equilibria, linear and nonlinear stability of the classical kind of relative equilibria of this generalized problem are investigated in the framework of geometric mechanics.

The conditions of linear stability of the relative equilibria are obtained based on the characteristic equation of the linear system matrix at the relative equilibria, which is given through the multiplication of the Poisson tensor and Hessian matrix of the variational Lagrangian. The conditions of nonlinear stability of the relative equilibria are derived with the energy-Casimir method through the distribution of the eigenvalues of the projected Hessian matrix of the variational Lagrangian.

With the stability conditions, both the linear and nonlinear stability of the relative equilibria are investigated in a wide range of the parameters of the gravity field and the rigid body by using the numerical method. The stability region is plotted on the plane of the mass distribution parameters of the rigid body in the cases of different values of the zonal harmonic $J_2$ and the characteristic dimension of the rigid body.

Similar to the classical attitude stability in a central gravity field, the linear stability region is consisted of two regions located in the first and third quadrant of the $\sigma_y - \sigma_x$ plane respectively, which are analogues of the Lagrange region and the DeBra-Delp region respectively. The nonlinear stability region is the subset of the linear stability region in the first quadrant, the analogue of the Lagrange region.

Both the zonal harmonic $J_2$ and the characteristic dimension of the rigid body have significant influences on the linear and nonlinear stability. When the characteristic dimension of the rigid body is small, the analogue of the Lagrange region in the first quadrant of the $\sigma_y - \sigma_x$ plane is actually the Lagrange region. When the characteristic dimension of the rigid body is large enough, the analogue of the Lagrange region is reduced by a triangle and this triangle expands as the zonal harmonic $J_2$ increases. For a given value of the characteristic dimension of the rigid body, as the zonal harmonic $J_2$ increases, the analogue of the DeBra-Delp region in the third quadrant of the $\sigma_y - \sigma_x$ plane expands in the direction of the boundary of the DeBra-Delp region, and cross the boundary of the DeBra-Delp region at $J_2 = 0$.

Our results on the stability of the relative equilibria are very useful for the studies on the motion of many natural satellites in our solar system, whose motion are close to the relative equilibria.

## Appendix: Formulations of Coefficients in Characteristic Equations

The explicit formulations of the coefficients in the characteristic equations Eqs. (38) and (50) are given as follows:

$$A_2 = -\frac{m}{2R_e^5}\left(3I_{yy}I_{zz}\mu + 12R_e^2 mI_{xx}\mu + 9mI_{zz}\mu\varepsilon - R_e^2 mI_{zz}\mu + 9I_{zz}^2\mu - 3R_e^2 m^2\mu\varepsilon \right.$$
$$\left. -12I_{xx}\mu I_{zz} - 4R_e^5 m\Omega_e^2 I_{zz} + 2m^2 R_e^7 \Omega_e^2 - 9R_e^2 mI_{yy}\mu - 2R_e^4 m^2\mu\right), \quad (A.1)$$

$$A_0 = -\frac{1}{2R_e^{10}}\left(-9\mu I_{yy} + 12I_{xx}\mu - 2m\mu R_e^2 + 2mR_e^5\Omega_e^2 - 3m\mu\varepsilon - 3\mu I_{zz}\right)*$$
$$\left(3m^2 R_e^7 \Omega_e^2 - 2R_e^4 m^2 \mu - 6R_e^2 m^2 \mu\varepsilon - R_e^5 m\Omega_e^2 I_{zz} - 8R_e^2 mI_{zz}\mu - 6R_e^2 mI_{yy}\mu \right.$$
$$\left. +12R_e^2 mI_{xx}\mu - 6mI_{zz}\mu\varepsilon - 6I_{yy}I_{zz}\mu + 12I_{xx}\mu I_{zz} - 6I_{zz}^2\mu\right), \quad (A.2)$$

$$B_4 = -\frac{1}{2R_e^5}\left(-5R_e^2 I_{yy} mI_{xx}\mu - 3I_{yy}^2 I_{xx}\mu + 2R_e^5 mI_{xx} I_{zz}\Omega_e^2 + 12I_{yy}I_{xx}^2\mu - 2R_e^4 m^2 I_{xx}\mu \right.$$
$$-3R_e^2 m^2 I_{xx}\mu\varepsilon - 4R_e^5 I_{yy} mI_{xx}\Omega_e^2 + 2R_e^7 m^2 I_{xx}\Omega_e^2 + 2R_e^5 I_{yy} m\Omega_e^2 I_{zz}$$
$$\left. -9I_{yy} mI_{xx}\mu\varepsilon - 9I_{yy}I_{xx} I_{zz}\mu - 2R_e^5 m\Omega_e^2 I_{zz}^2 + 12R_e^2 mI_{xx}^2\mu - 9R_e^2 mI_{xx} I_{zz}\mu\right), \quad (A.3)$$

$$B_2 = \frac{1}{2R_e^8}\left(-8I_{yy} mR_e^5 \Omega_e^2 I_{zz}\mu + 27I_{yy} R_e^3 I_{xx} I_{zz}\mu\Omega_e^2 + 2m^2 R_e^7 I_{xx}\Omega_e^2\mu - 5mR_e^5 I_{xx} I_{zz}\mu\Omega_e^2 \right.$$
$$+3m^2 R_e^5 \Omega_e^2 I_{zz}\mu\varepsilon - 2m^2 R_e^{10}\Omega_e^4 I_{zz} + 6R_e^2 m^2 I_{xx}\mu^2\varepsilon + 19I_{yy} mR_e^5 I_{xx}\Omega_e^2\mu$$
$$+27mI_{xx}\mu^2\varepsilon I_{zz} + 11mR_e^5\Omega_e^2 I_{zz}^2\mu + 2m^2 R_e^7\Omega_e^2 I_{zz}\mu - 3I_{yy}^2 mR_e^5\Omega_e^2\mu - 2I_{yy} m^2 R_e^7\Omega_e^2\mu$$
$$-2m^2 R_e^{10} I_{xx}\Omega_e^4 + 2I_{yy} m^2 R_e^{10}\Omega_e^4 + 9R_e^3\Omega_e^2 I_{zz}^3\mu - 36mI_{xx}^2\mu^2\varepsilon + 9\mu^2 m^2\varepsilon^2 I_{xx}$$
$$+2mR_e^8\Omega_e^4 I_{zz}^2 - 6I_{yy} R_e^3\Omega_e^2 I_{zz}^2\mu - 9I_{yy} mR_e^3\Omega_e^2 I_{zz}\mu\varepsilon - 9mR_e^3 I_{xx}\Omega_e^2 I_{zz}\mu\varepsilon$$
$$-21R_e^3 I_{xx} I_{zz}^2\mu\Omega_e^2 - 2mR_e^8 I_{xx}\Omega_e^4 I_{zz} - 3I_{yy} m^2 R_e^5\Omega_e^2\mu\varepsilon - 3m^2 R_e^5 I_{xx}\Omega_e^2\mu\varepsilon$$
$$-24I_{yy} R_e^3 I_{xx}^2\Omega_e^2\mu - 3I_{yy}^2 R_e^3\Omega_e^2 I_{zz}\mu - 12mR_e^5 I_{xx}^2\Omega_e^2\mu + 6I_{yy}^2 R_e^3 I_{xx}\Omega_e^2\mu$$
$$+9mR_e^3\Omega_e^2 I_{zz}^2\mu\varepsilon + 2I_{yy} mR_e^8 I_{xx}\Omega_e^4 + 18I_{yy} mR_e^3 I_{xx}\Omega_e^2\mu\varepsilon + 9I_{yy} mI_{xx}\mu^2\varepsilon$$
$$\left. +12R_e^3 I_{xx}^2 I_{zz}\mu\Omega_e^2 - 2I_{yy} mR_e^8\Omega_e^4 I_{zz}\right), \quad (A.4)$$

$$B_0 = \frac{\Omega_e^2}{2R_e^8}(I_{yy} - I_{zz})(-9R_e^3 I_{zz}^2 \mu \Omega_e^2 - 11I_{zz} mR_e^5 \Omega_e^2 \mu - 3I_{yy} R_e^3 I_{zz} \mu \Omega_e^2 + 21I_{zz} R_e^3 I_{xx} \mu \Omega_e^2$$
$$-9I_{zz} mR_e^3 \Omega_e^2 \mu \varepsilon - 27I_{zz} m\mu^2 \varepsilon + 14mR_e^5 I_{xx} \mu \Omega_e^2 - 9I_{yy} m\mu^2 \varepsilon + 36mI_{xx} \mu^2 \varepsilon$$
$$-6R_e^2 m^2 \mu^2 \varepsilon - 2m^2 R_e^7 \Omega_e^2 \mu - 3I_{yy} mR_e^5 \Omega_e^2 \mu - 12R_e^3 I_{xx}^2 \mu \Omega_e^2 + 3I_{yy} R_e^3 I_{xx} \mu \Omega_e^2$$
$$-9\mu^2 m^2 \varepsilon^2 + 9mR_e^3 I_{xx} \Omega_e^2 \mu \varepsilon + 3m^2 R_e^5 \Omega_e^2 \mu \varepsilon + 2m^2 R_e^{10} \Omega_e^4), \tag{A.5}$$

$$C_2 = -\frac{1}{mI_{yy}}\left(I_{yy} + m^2 \Omega_e^2 R_e^2 I_{yy} + m\Omega_e^2 I_{zz} I_{yy} + m\right), \tag{A.6}$$

$$C_1 = \frac{1}{mI_{yy}}\left(\Omega_e^2 I_{zz} I_{yy} + m^2 \Omega_e^2 R_e^2 + 1 + m\Omega_e^2 I_{zz} - m\Omega_e^2 I_{yy}\right), \tag{A.7}$$

$$C_0 = \frac{\Omega_e^2}{mI_{yy}}(I_{yy} - I_{zz}), \tag{A.8}$$

$$D_2 = -\frac{1}{2R_e^5 I_{xx}}(2mI_{xx} \Omega_e^2 R_e^7 + 2R_e^5 I_{xx} \Omega_e^2 I_{zz} + 2R_e^5 + 2mR_e^2 \mu I_{xx} + 6R_e^2 m\varepsilon \mu I_{xx}$$
$$+9m\varepsilon \mu I_{xx} - 12\mu I_{xx}^2 + 9I_{zz} \mu I_{xx} + 3\mu I_{yy} I_{xx}), \tag{A.9}$$

$$D_1 = \frac{1}{2R_e^8 I_{xx}}(27m\varepsilon \mu^2 I_{xx} I_{zz} + 11mR_e^5 \mu I_{xx} \Omega_e^2 I_{zz} - 2R_e^8 I_{xx} \Omega_e^2 + 3\mu I_{yy} R_e^3 \Omega_e^2 I_{zz} I_{xx}$$
$$+2mR_e^5 \mu + 9mR_e^3 \varepsilon \mu - 2m^2 R_e^{10} I_{xx} \Omega_e^4 + 9R_e^3 \mu I_{xx} \Omega_e^2 I_{zz}^2 + 9\mu^2 I_{yy} mI_{xx} \varepsilon$$
$$-3m^2 R_e^5 \varepsilon \mu I_{xx} \Omega_e^2 + 2mR_e^{10} \Omega_e^2 - 36m\varepsilon \mu^2 I_{xx}^2 + 3\mu I_{yy} m\Omega_e^2 I_{xx} R_e^5 + 6mR_e^5 \varepsilon \mu$$
$$-12mR_e^5 \mu I_{xx}^2 \Omega_e^2 + 9m^2 \varepsilon^2 \mu^2 I_{xx} + 2R_e^8 \Omega_e^2 I_{zz} + 3\mu I_{yy} R_e^3 - 12R_e^3 \mu I_{xx}$$
$$+2m^2 R_e^7 \mu I_{xx} \Omega_e^2 - 12R_e^3 \mu I_{xx}^2 \Omega_e^2 I_{zz} + 6R_e^2 m^2 \varepsilon \mu^2 I_{xx} + 9R_e^3 I_{zz} \mu$$
$$+9mR_e^3 \varepsilon \mu I_{xx} \Omega_e^2 I_{zz}), \tag{A.10}$$

$$D_0 = \frac{1}{2R_e^8 I_{xx}}(2m^2 R_e^{10} \Omega_e^4 + 36m\varepsilon \mu^2 I_{xx} - 9m^2 \varepsilon^2 \mu^2 - 3\mu I_{yy} \Omega_e^2 R_e^5 m - 6R_e^2 m^2 \varepsilon \mu^2$$
$$+21R_e^3 \mu I_{xx} \Omega_e^2 I_{zz} - 11mR_e^5 \mu \Omega_e^2 I_{zz} - 9R_e^3 \mu \Omega_e^2 I_{zz}^2 + 9mR_e^3 \varepsilon \mu I_{xx} \Omega_e^2$$
$$-2m^2 R_e^7 \mu \Omega_e^2 - 27m\varepsilon \mu^2 I_{zz} + 3\mu I_{yy} R_e^3 \Omega_e^2 I_{xx} - 9\mu^2 I_{yy} m\varepsilon + 14mR_e^5 \mu I_{xx} \Omega_e^2$$
$$+3m^2 R_e^5 \varepsilon \mu \Omega_e^2 - 3\mu I_{yy} R_e^3 \Omega_e^2 I_{zz} - 12R_e^3 \mu I_{xx}^2 \Omega_e^2 - 9mR_e^3 \varepsilon \mu \Omega_e^2 I_{zz}), \tag{A.11}$$

$$E_1 = \frac{1}{mR_e^5 I_{zz}(1 + R_e^2 + m^2 R_e^2 \Omega_e^2)}(6R_e^2 m^2 \varepsilon \mu I_{zz} + 2R_e^2 m^2 I_{zz} \mu - R_e^5 I_{zz} - 12mR_e^2 I_{zz} \mu I_{xx}$$
$$+6mI_{zz}^2 \mu - mR_e^7 - 3m^2 R_e^7 \Omega_e^2 I_{zz} + 6\mu I_{yy} R_e^2 I_{zz} m - 12mI_{zz} \mu I_{xx} - R_e^7 \Omega_e^2 m^3$$
$$+2m^2 R_e^4 I_{zz} \mu + 6\mu I_{yy} I_{zz} m + 6mR_e^2 I_{zz}^2 \mu + 6m^2 \varepsilon \mu I_{zz}), \tag{A.12}$$

$$E_0 = \frac{1}{mR_e^5 I_{zz}(1 + R_e^2 + m^2 R_e^2 \Omega_e^2)}(-6\mu I_{yy} I_{zz} - 2m^2 R_e^4 \mu + 3m^2 R_e^7 \Omega_e^2 - 8\mu R_e^2 I_{zz} m$$
$$+12mR_e^2 \mu I_{xx} + 12I_{zz} \mu I_{xx} - 6\mu I_{yy} mR_e^2 - 6m\varepsilon \mu I_{zz} - 6I_{zz}^2 \mu - 6R_e^2 m^2 \varepsilon \mu$$
$$-mR_e^5 \Omega_e^2 I_{zz}), \tag{A.13}$$

$$F_1 = \frac{1}{2mR_e^5}\left(-3\mu I_{zz}m + 12m\mu I_{xx} - 9\mu I_{yy}m - 2R_e^5 - 3m^2\varepsilon\mu - 2R_e^2 m^2\mu\right), \quad (A.14)$$

$$F_0 = \frac{1}{2mR_e^5}\left(-2\Omega_e^2 R_e^5 m - 12\mu I_{xx} + 3\mu m\varepsilon + 9\mu I_{yy} + 3\mu I_{zz} + 2\mu mR_e^2\right). \quad (A.15)$$

## Acknowledgements

This work is supported by the Innovation Foundation of BUAA for PhD Graduates.

## References

Aboelnaga, M.Z., Barkin, Y.V.: Stationary motion of a rigid body in the attraction field of a sphere. Astronom. Zh. **56**(3), 881–886 (1979)

Balsas, M.C., Jiménez, E.S., Vera, J.A.: The motion of a gyrostat in a central gravitational field: phase portraits of an integrable case. J. Nonlinear Math. Phy. **15**(s3), 53–64 (2008)

Barkin, Y.V.: Poincaré periodic solutions of the third kind in the problem of the translational-rotational motion of a rigid body in the gravitational field of a sphere. Astronom. Zh. **56**, 632–640 (1979)

Beck, J.A., Hall, C.D.: Relative equilibria of a rigid satellite in a circular Keplerian orbit. J. Astronaut. Sci. **40**(3), 215–247 (1998)

Beletskii, V.V., Ponomareva, O.N.: A parametric analysis of relative equilibrium stability in a gravitational field. Kosm. Issled. **28**(5), 664–675 (1990)

Bellerose, J., Scheeres, D.J.: Energy and stability in the full two body problem. Celest. Mech. Dyn. Astron. **100**, 63–91 (2008)

Boué, G., Laskar, J.: Spin axis evolution of two interacting bodies. Icarus **201**, 750–767 (2009)

Breiter, S., Melendo, B., Bartczak, P., Wytrzyszczak, I.: Synchronous motion in the Kinoshita problem. Applications to satellites and binary asteroids. Astron. Astrophys. **437**(2), 753–764 (2005)

Broucke, R.A.: Numerical integration of periodic orbits in the main problem of artificial satellite theory. Celest. Mech. Dyn. Astron. **58**, 99–123 (1994)

Hall, C.D.: Attitude dynamics of orbiting gyrostats, in: Prętka-Ziomek, H., Wnuk, E., Seidelmann, P.K., Richardson, D. (Eds.), Dynamics of Natural and Artificial Celestial Bodies. Kluwer Academic Publishers, Dordrecht, pp. 177–186 (2001)

Hughes, P.C.: Spacecraft Attitude Dynamics, John Wiley, New York, pp. 281–298 (1986)

Kinoshita, H.: Stationary motions of an axisymmetric body around a spherical body and their stability. Publ. Astron. Soc. Jpn. **22**, 383–403 (1970)

Koon, W.-S., Marsden, J.E., Ross, S.D., Lo, M., Scheeres, D.J.: Geometric mechanics and the dynamics of asteroid pairs. Ann. N. Y. Acad. Sci. **1017**, 11–38 (2004)

Maciejewski, A.J.: Reduction, relative equilibria and potential in the two rigid bodies problem. Celest. Mech. Dyn. Astron. **63**, 1–28 (1995)

Marsden, J.E., Ratiu, T.S.: Introduction to Mechanics and Symmetry, TAM Series 17, Springer Verlag, New York (1999)

McMahon, J.W., Scheeres, D.J.: Dynamic limits on planar libration-orbit coupling around an oblate primary. Celest. Mech. Dyn. Astron. doi: 10.1007/s10569-012-9469-0 (in press)

Scheeres, D.J.: Stability in the full two-body problem. Celest. Mech. Dyn. Astron. **83**, 155–169 (2002)

Scheeres, D.J.: Stability of relative equilibria in the full two-body problem. Ann. N. Y. Acad. Sci.

**1017**, 81–94 (2004)

Scheeres, D.J.: Spacecraft at small NEO. arXiv: physics/0608158v1 (2006)

Scheeres, D.J.: Stability of the planar full 2-body problem. Celest. Mech. Dyn. Astron. **104**, 103–128 (2009)

Teixidó Román, M.: Hamiltonian Methods in Stability and Bifurcations Problems for Artificial Satellite Dynamics. Master Thesis, Facultat de Matemàtiques i Estadística, Universitat Politècnica de Catalunya, pp. 51–72 (2010)

Vereshchagin, M., Maciejewski, A.J., Goździewski, K.: Relative equilibria in the unrestricted problem of a sphere and symmetric rigid body. Mon. Not. R. Astron. Soc. **403**, 848–858 (2010)

Wang, Y., Xu, S.: Gravitational orbit-rotation coupling of a rigid satellite around a spheroid planet. J. Aerosp. Eng. doi: 10.1061/(ASCE)AS.1943-5525.0000222 (in press)

Wang, Y., Xu, S.: Hamiltonian structures of dynamics of a gyrostat in a gravitational field. Nonlinear Dyn. **70**(1), 231–247 (2012)

Wang, Y., Xu, S.: Symmetry, reduction and relative equilibria of a rigid body in the $J_2$ problem. Adv. Space Res. **51**(7), 1096–1109 (2013a)

Wang, Y., Xu, S.: Gravity gradient torque of spacecraft orbiting asteroids. Aircr. Eng. Aerosp. Tec. **85**(1), 72–81 (2013b)

Wang, L.-S., Krishnaprasad, P.S., Maddocks, J.H.: Hamiltonian dynamics of a rigid body in a central gravitational field. Celest. Mech. Dyn. Astron. **50**, 349–386 (1991)

Wang, L.-S., Maddocks, J.H., Krishnaprasad, P.S.: Steady rigid-body motions in a central gravitational field. J. Astronaut. Sci. **40**(4), 449–478 (1992)

Wang, L.-S., Lian, K.-Y., Chen, P.-T.: Steady motions of gyrostat satellites and their stability. IEEE T. Automat. Contr. **40**(10), 1732–1743 (1995)

Wisdom, J.: Rotational dynamics of irregularly shaped natural satellites. Astron. J. **94**, 1350–1360 (1987)